\documentclass{aa}  

\usepackage{graphicx}
\usepackage{xspace}
\usepackage[]{hyperref}
\usepackage{lineno}

\usepackage[nameinlink,noabbrev]{cleveref}
\Crefname{equation}{Eq.}{Eqs.}
\Crefname{eqnarray}{Eq.}{Eqs.}
\Crefname{section}{Sect.}{Sects.}
\Crefname{figure}{Fig.}{Figs.}
\crefname{equation}{Equation}{Equations}
\crefname{section}{Section}{Sections}
\crefname{figure}{Figure}{Figures}
\creflabelformat{equation}{#2#1#3}

\newcommand{\xmm}{\emph{XMM-Newton}\xspace}
\newcommand{\nustar}{\emph{NuSTAR}\xspace}

\newcommand{\kms}{\mathrm{km\,s^{-1}}}
\newcommand{\Msun}{\mathrm{M_{\odot}}}

\newcommand{\vout}{v_{\mathrm{out}}}

\newcommand{\NH}{N_{\mathrm{H}}}

\hypersetup{
    colorlinks=true,
    linkcolor=blue,
    filecolor=magenta,      
    urlcolor=cyan,
    citecolor=blue
}

\usepackage{txfonts}
\usepackage{natbib}

\usepackage[toc,page]{appendix}

\defcitealias{Gianolli2024}{G24}
\defcitealias{Tombesi10}{T10}
\defcitealias{Matzeu23}{M23}
\defcitealias{Chartas21}{C21}

\begin{document} 

   \title{Discovery of ultra-fast outflows with v$_{\rm out}>0.3 \rm c$ in local bright active galactic nuclei}

   \author{L. Borrelli
          \inst{1, 2}
          \and
          M. Guainazzi
          \inst{3}
          \and
          G. Lanzuisi
          \inst{2}
          \and
          E. Piconcelli
          \inst{4}
          \and 
          L. Pentericci
          \inst{4}
          \and
          A. Luminari
          \inst{4,5}
          \and
          J. Svoboda
          \inst{6}
          \and
          A. Borkar
          \inst{6}
          }

   \institute{
                Dipartimento di Fisica e Astronomia "Augusto Righi", Università di Bologna, via Gobetti 93/2, 40129 Bologna, Italy 
                \and
                INAF –  OAS, Osservatorio di Astrofisica e Scienza dello Spazio Bologna, Via Piero Gobetti, 93/3, 40129 Bologna, Italy
                \and
                European Space Agency (ESA), European Space Research and Technology Centre (ESTEC), Keplerlaan 1, 2201 AZ Noordwijk, The Netherlands
                \and
                INAF – Osservatorio Astronomico di Roma, Via Frascati 33, I-00040, Monte Porzio Catone, Italy
                \and
                INAF – Istituto di Astrofisica e Planetologia Spaziali, Via del Fosso del Caveliere 100, I-00133 Roma, Italy
                \and
                Astronomical Institute of the Czech Academy of Sciences, Boční II 1401/1, 14100 Praha 4, Czech Republic
             }

   \date{Received ; accepted }

  \abstract
   {Ultra-fast outflows (UFOs) are mildly relativistic (outflow velocity $\vout>0.1\rm c$) nuclear winds detected as blueshifted absorption lines from highly ionized, dense gas in the X-ray spectra of active galactic nuclei (AGNs). The AGN feedback mechanism is believed to be powered by these outflows, which can inject a large amount of energy and momentum into the surrounding interstellar medium, shaping the coevolution of the AGNs and their host galaxies.}
   {We performed a systematic search and rigorous statistical assessment of the presence of UFOs in the $7-12$ keV band, in a sample of bright local AGNs. This study also aims to understand whether the presence and characteristics of UFOs depend on the state of the sources, by studying the relations between the incidence of UFOs and the accretion properties of AGNs.}
   {We collected X-ray spectroscopic flux-limited \xmm data of 33 observations of local ($z<0.2$) type 1 AGNs.
   We modeled their spectra in the $2-12$ keV band using a combination of direct-continuum and reflection components and searched for absorption features. This represents the first systematic search for UFOs up to 12 keV. We performed Monte Carlo (MC) simulations to assess the statistical significance of the detected lines. 
   }
   {We report strong detections of UFOs in six sources of the sample at the $\geq 95\%$ confidence level via MC simulations, corresponding to a fraction of $\sim18\%$ in our sample. From the observed energies of each absorption line, we evaluated the respective wind velocities, which in some cases exceed 40\% of the speed of light. The velocity distribution found in this work is therefore shifted to higher energies than those found in previous searches for UFOs in local sources, which were limited to 10 keV. Moreover, our analysis shows no correlation between the accretion properties of the super massive black holes (SMBHs) and the presence of winds. Finally, the comparison with previous studies of nuclear winds in local AGNs highlights the temporal variability of UFOs.
   }
  {}


    \maketitle
\nolinenumbers 
\section{Introduction} \label{intro}
Active galactic nuclei (AGNs) are powered by the accretion of gas onto super massive black holes (SMBHs). AGNs release huge amounts of radiation that can strongly influence the nuclear environment, the SMBHs' growth, and the evolution of the host galaxy, as witnessed by scaling relations between the SMBH mass and galactic properties \citep{Kormendy95,Magorrian98,Ferrarese00,Gebhardt00,Kormendy13}. AGN feedback, which is believed to be driven by powerful outflows from the accretion disk \citep{Silk98, Zubova12}, could be a promising mechanism to account for some of the most significant unresolved issues in astronomy, such as the underlying reasons for the strong correlation between SMBH mass and the stellar bulge velocity dispersion of the host galaxy (the $M_{BH}-\sigma$ relation) and the evolution of the quasar luminosity function over cosmic timescales across different wavelengths. 

\cite{Laha21} identify ionized outflows as an essential element of AGN feedback.
Ultra-fast outflows (UFOs) are highly ionized (log $\xi=3-6 $ $\rm erg~s^{-1}~cm $, with $\xi=\frac{L_{ion}}{n_H r^2}$ being the ionization parameter) outflows observed as strongly blueshifted absorption features in the hard X-ray band associated with iron (multiplet FeXXV and doublet FeXXVI, for which we assume centroid rest-frame energies at $E_{FeXXV}=6.7$ keV and $E_{FeXXVI}=6.97$ keV, respectively, which is sufficient at CCD resolution); they have mildly relativistic velocities up to $\vout=0.2-0.4\rm c$ \citep{Chartas02,ReevesBraito19,Laurenti21}.
They are now routinely observed in $\sim30\%$ of local Seyferts \citep[hereafter \citetalias{Tombesi10}]{Tombesi10} and quasars (QSOs) 
\citep[hereafter \citetalias{Matzeu23}]{Matzeu23}.

Due to their high velocity, high ionization, and extreme variability, these near-relativistic outflows are most likely launched from the inner part of the accretion disk in the deepest part of the SMBH gravitational potential. Variability on timescales as short as 1 day has been reported; therefore, the typical location has been estimated to be within 0.01-0.1 pc from the central SMBH \citep{Kraemer18,Fukumura22}. Some cases \citep{Chartas18IRAS13224,Giustini11} even report variability on intra-observation timescales ($10^4-10^5$ s), placing the origin of UFOs even closer to the nuclear region.

Strong correlations between the bolometric luminosity of the AGNs and both the kinetic power \citep{Fiore17} and outflow velocity \citep{Matzeu23,Xu2025_PDS} of UFOs have been detected. These correlations suggest that faster winds are preferentially driven in high-luminosity or high-accreting AGNs. While the origin of these ionized outflows is not fully understood, three primary mechanisms have been proposed: radiation driving, magnetic driving, and thermal driving. 

Determining which mechanism dominates is challenging. Although radiation likely plays a crucial role, it faces significant physical limitations. UFOs appear to consist of dust-free, highly ionized material; this lack of dust and opacity in UV or X-ray lines renders standard radiatively driven flow models inefficient. While radiation scattering on free electrons contributes \citep{Tombesi13}, it may be insufficient to accelerate the fastest winds \citep{Luminari20}. Consequently, magneto-hydrodynamical (MHD) effects in the accretion disk are increasingly viewed as a necessary component for generating these high-velocity outflows \citep{Laha21}.

To distinguish between these mechanisms and assess their impact on host galaxies, it is essential to systematically detect the fastest wind components. This is critical because the potential feedback impact of outflows scales steeply with velocity ($\dot{E}_{out}\propto \vout^3$) \citep{KingPounds15}. 
While individual cases of $\vout > 0.3\rm c$ UFOs have been reported in local AGNs through targeted, multi-observatory campaigns \citep{Reeves18A,Reeves18B,Luminari23} and such velocities have been  observed in high-z lensed QSOs where the features are redshifted into the standard band (\citet{Chartas02}, \citetalias{Chartas21}), a systematic exploration of the $10-12$ keV regime in a flux-limited sample of local sources has never been attempted. 

The specific aim of this work is to perform the first systematic, blind search for UFO absorption features in the $7-12$ keV rest-frame band in a flux-limited sample of bright local AGN observed with \xmm, employing a homogeneous approach in both spectral and statistical analysis. By extending beyond the 10 keV boundary adopted by all previous systematic UFO surveys (e.g., \citetalias{Tombesi10}, \citet{Gofford13,Igo20}), we aim to determine whether the extreme velocities ($\vout > 0.3\rm c$) inferred from individual source studies and high-z observations are a common property of local AGN winds.

We also aim to understand whether the presence of such UFOs depends on the accretion properties of the sources within our parent sample, which \citet{Svoboda17} characterized self-consistently using simultaneous UV and X-ray data, as is the case for nuclear winds in stellar-mass black hole binaries (BHBs). In fact, in the case of BHBs, strong connections between X-ray winds and the source’s accretion state have been found: winds appear mostly in sources in the soft state, characterized by a high accretion rate \citep{Ponti12}. 

The paper is organized as follows: Sec. \ref{Sect. 2} presents the aims of the study and shows the analyzed sample. In Sec. \ref{data reduction}, we describe the data reduction process. In Sec. \ref{Data Analysis}, we present the methodology for data analysis and evaluate the statistical relevance and properties of the detected absorption. In Sec. \ref{Sec. 5}, we present and discuss the most significant results of the analysis we performed. In Sec. \ref{conclusions}, a summary of the results is provided. Throughout this paper, the following cosmological constants are assumed: $H_0 = 70~\kms~\rm Mpc^{-1}$, $\Omega_\Lambda = 0.73$, and $\Omega_M = 0.27$.

\section{Sample selection} 
\label{Sect. 2}

We opted for the \cite{Svoboda17} catalog updated with observations up to 2024, which contains numerous sources for which the UV and X-ray emission have been measured through simultaneous coverage by the OM and EPIC instruments on board \xmm. Since a high number of spectral counts is required to detect UFOs, the following selection criterion was applied to the catalog to obtain our sample: $F_{2-12\rm keV}\geq1.5\times10^{-11}~ \rm erg~s^{-1}~cm^{-2}$.

\begin{figure}[t]
    \centering
    \includegraphics[scale=1, angle=0, width=8cm,height=8cm,keepaspectratio]{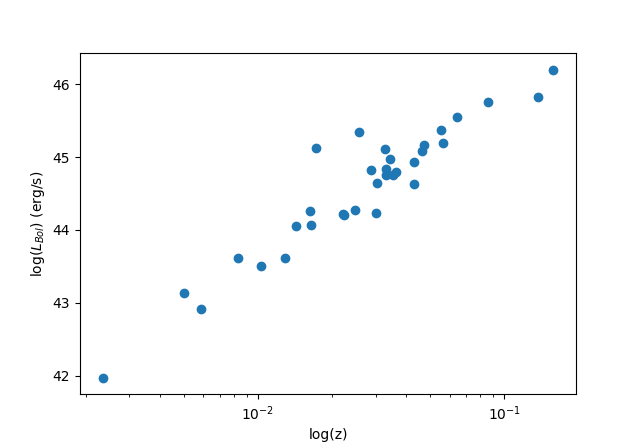}
    \caption{Bolometric luminosity of the 33 AGNs considered in the sample (see Table \ref{Table 4.1}) plotted against their redshift.
    }
    \label{L vs z}
\end{figure}

\cite{Svoboda17} draw an analogy between stellar mass black holes in different spectral states and populations of AGNs. They show that AGNs occupy analogous regions of the hardness-flux (or accretion) diagram when the AGN energy output is measured in a way that separates the thermal disk from the hard X-ray corona. For each source, the bolometric output is decomposed into disk luminosity and hot corona emission. The redshift and Galactic extinction corrected UV flux were used to estimate the disk luminosity $L_D$ at $ 2910 \r{A}$ (see \citealt{Svoboda17} for details), which can be defined as
$$L_D=A\cdot4\pi D_L^2 \lambda F_{\lambda, 2910 \r{A}},$$
where $D_L$ is the luminosity distance constrained from the redshift measurement and $A$ is an arbitrary factor, which can be chosen so that the sum of the disk and power-law luminosity, $L_{tot}$, roughly corresponds to the bolometric luminosity $L_{Bol}$. The corona (power-law) luminosity ($L_P$) can be defined from the X-ray luminosity by extrapolating it to the energy interval $0.1-100$ keV
$$L_P=4\pi D_L^2 F_{0.1-100 \rm keV},$$
where $F_{0.1-100 \rm keV}$ is the X-ray flux at the $0.1-100$ keV energy range calculated by an extrapolation of the observed 2–12 keV flux.  In the end, we can assume $L_{tot}\simeq L_{Bol}=L_D+L_P$. 
The spectral hardness parameter H is defined as: $$H=\frac{L_P}{L_P+L_D}.$$

This highlights the importance of having simultaneous measures in UV and X-ray bands: not only does UV data lead to a precise estimation of the accretion-disk luminosity, but is also crucial both to reduce the effect of variability in the measurements of the disk and power-law luminosities, and to constrain the exact value of H, and thus evaluate the ionization state of the source. The results of \cite{Svoboda17} show that radio loudness increases on average with H, and with decreasing Eddington Ratio ($\lambda_{Edd}=\frac{L_{Bol}}{L_{Edd}}$, that is a measure of the accretion rate of the SMBH), suggesting that the jet emission occurs in the low accretion states, consistent with BHBs \citep{moravec22}.
The original catalog contained multiple observations for several sources; we decided to analyze one observation per AGN. The longest observation for each source in the \xmm public archive was selected to ensure good spectral statistical quality while avoiding source-dependent bias in the sample. This led to the analysis of 33 \xmm observations for the AGN listed in Table \ref{Table 4.1}, while Table \ref{Table 4.2} summarizes the details of the selected observations. Figure \ref{L vs z} shows the distribution of $L_{Bol}$ versus redshift for the sources of our sample. 

$L_{Bol}$ and the H information are taken from the \citep{Svoboda17} catalog. 
We used more recent observations than those from 2017 in 9 cases (out of 33) where significantly longer exposures were available.
We observe the typical long-term X-ray variability of up to 50\%. However, given the relative contribution of $L_D$ and $L_P$, the impact on $L_{Bol}$ and $H$ is limited (up to $\sim15\%$), therefore, we keep the values from \cite{Svoboda17}.

For the masses of the selected black holes, we report the values inferred by recent studies using the reverberation mapping method or the single-epoch method, based on Balmer line properties, in Table \ref{Table 4.1}. Finally, from $L_{Bol}$ and BH masses, we calculated the Eddington ratios. 

\section{Data reduction} 
\label{data reduction}
The \xmm data reduction was carried out with the Science Analysis System (SAS)
software package 21.0.0, specifically designed to reduce and analyze data collected by the \xmm observatory \citep{Schartel_2022}. 
In this work, we have studied only the spectra extracted from the EPIC-pn camera, as it has the highest collecting area, especially in the hard band.
We started from the EPIC observation data files (ODFs) available in the \xmm archive, and performed standard data reduction. We removed noise events and periods dominated by soft proton flares, extracting a single event, high-energy (10-12 keV) light curve to identify intervals of flaring background; a threshold in counts/seconds was chosen to define the "low background" intervals (0.4 for EPIC-pn). Then, we obtained the net exposure time (see Table \ref{Table 4.2}) for each observation, representing the time available for analysis. For most of them, around 60-70\% of the nominal duration was usable.

\begin{figure}[ht]
    \centering
    \includegraphics[scale=1, angle=0, width=8cm,keepaspectratio]{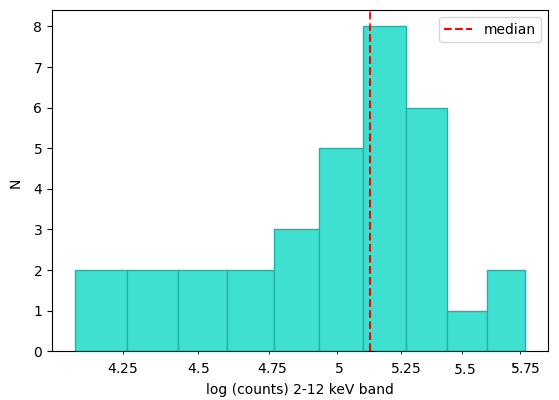}
    \caption{Distribution of the net counts in the 2-12 keV band for the pn camera (see Table \ref{Table 4.2}) after removing periods dominated by soft proton flares.}
    \label{counts}
\end{figure}

During the creation of the spectra, we visually checked if the observations were affected by pile-up, which can occur for bright sources ($F_{0.2-12 \rm keV} \gtrsim 10^{-11}$ erg s$^{-1}$ cm$^{-2}$ for EPIC pn, depending on the observation mode). To overcome the pile-up in SBS~1301+540, whose observations were taken in full-window mode, we repeated the extraction process, excluding the PSF center (see \citealt{Schartel_2022}) over an area of radius 200-250 pixels and verifying that pile-up was negligible in that case.
The distribution of the net counts collected in the 2-12 keV band (see Table \ref{Table 4.2}) for the pn camera is shown in Fig. \ref{counts}, with a median value of $\sim1.4\times10^5$ counts. We then extracted the spectra for source and background, and finally created the RMF and ARF ("response matrices") using the standard SAS tools {\sc arfgen} and {\sc rmfgen}.  

As we extend the spectral band for the analysis to 12 keV, a regime where the effective area of the \xmm telescopes decreases significantly, calibration uncertainties increase, and particle background dominates the overall instrument background, we applied three different prescriptions to estimate the background spectrum:

\begin{itemize}

    \item the observation background: we extracted spectra from a spatial
    region close to the source extraction region on the same pn CCD, and at
    the same approximate distance from the readout node in pn to ensure that similar corrections for Charge Transfer Efficiency apply. Since the vast majority of the observations were taken in small window mode, this standard technique was considered not sufficient alone, because the small chip area used in this mode is almost entirely dominated by the source emission;

    \item the blank sky (BS) background: we extracted background spectra
     from black field event lists \citep{carter07} for the full window mode observation, using the same spatial region in detector coordinates as the source;

    \item the filter wheel closed (FWC) background: we extracted spectra from exposures where a filter is used to occult the optical path to the sky, available for all pn observing modes.
    These exposures enable a direct estimate of the instrumental unfocused particle background. As in the case of the blank sky background, spectra were extracted using the same spatial region in detector coordinates as the source.

\end{itemize}

All the sources had their spectra extracted first with the observational background method. The spectra that hinted at the presence of a nuclear wind were then re-extracted with both the BS background, where possible, and with the FWC background in all cases. \\
To properly bin the spectra, we adopted the optimal binning method outlined by \citet{KaastraBleeker16} (KB), which accounts for source and background levels, as well as instrument resolution. This scheme, implemented via \texttt{ftgrouppha} in \texttt{heasoft}\footnote{\url{https://heasarc.gsfc.nasa.gov/docs/software/lheasoft/}}, provides a variable bin size matched to the energy-dependent resolution of the EPIC-pn CCD and the spectral SN, and is specifically designed to maximize the information content for narrow, unresolved spectral features. \citetalias{Matzeu23} systematically compared this method against three other commonly used binning prescriptions (grpmin1, SN5, and OS3grp20) in the context of Fe K absorption line searches in \xmm spectra of the SUBWAYS sample. They found that the KB and grpmin1 binning produce nearly identical results in terms of detection rate and line significance, with KB making the narrow features more easily identifiable in the  spectra, while coarser schemes can dilute faint absorption features, making the KB binning the optimal choice for this type of analysis.\\
Since the KB optimal binning does not impose a minimum number of counts per bin, some spectral bins may contain too few counts for the $\chi^2$ approximation to be valid. We therefore adopted the Cash Statistic or $C_{stat}$ \citep{Cash79} throughout the spectral analysis, which provides unbiased parameter estimates and confidence intervals in the Poissonian regime, as demonstrated by \citet{Humphrey09} and \citet{Kaastra_2017}, even when the number of counts per bin is low.

\section{Data analysis}  \label{Data Analysis}
In this chapter, we outline the methodology used to fit the observed spectra, to search for UFO absorption features, and to assess the significance of the detected winds. 

\subsection{Continuum modeling} \label{Continuum modeling}
To model the 2-12 keV rest-frame spectra, we used a Baseline Model (BM) made of a combination of emission components due to direct continuum and reflection characteristic of AGN emission, as a starting point to reproduce the standard features present in AGNs, which we can write as: $\textsc{zTBabs*(MYTorus\_S+MYTorus\_L+zpow})$.
The components of the BM are the following: the power-law represents the continuum generated by Inverse Compton scattering of UV photons from the accretion disk and semi-relativistic electrons in the hot corona. \textsc{zTbabs} is a multiplicative component at the source redshift, used to reproduce the effect of cold absorption due to the material along the line of sight to the observer. The Galactic column density of the Milky Way was not considered since we are fitting at energies $>2$ keV where its contribution is negligible for our sources. \\
\textsc{MYTorus} \citep{Murphy09} is a model that reproduces the effects of reflection by the torus of the primary power-law emission, that is the Compton-scattered continuum (MYTorus\_S) plus the associated fluorescent lines (MYTorus\_L). The inclination angle and relative intensity are fixed to their default values, while the column density is set to be the same for the continuum and fluorescent lines components (coupled mode), with typical best-fit values around $N_H\sim10^{23}$~cm$^{-2}$.

\begin{table*}[h!]
\caption{Properties of the absorption features detected with high significance, as described in Sect. \ref{Statistical analysis and robustness of results}.}
\centering
  \renewcommand{\arraystretch}{1.25}
\begin{tabular}{l c c c c c c c c}
\hline
Source & $\Gamma$ & Energy & Width & |Depth| & $\Delta$ Cstat & Significance & |EQW| & v/c \\
\hline
(1) & (2) & (3) & (4) & (5) & (6) & (7) & (8) & (9)\\
\hline
\hline

Mkn 766* & $2.11_{-0.02}^{+0.01}$ & $9.46\pm0.32$ & $0.37\pm0.34$ & $10.98\pm0.79$ & 13.05 & 99.70\% & $133.6^{+28.9}_{-24.5}$  & $0.30\pm0.02$\\
   
Mrk 110* & $1.70_{-0.02}^{+0.04}$ & $8.18\pm0.18$ & $\le0.194$ & $8.71\pm0.31$ & 7.83 & 96.40\% & $39.7^{+7.4}_{-7.4}$  & $0.16\pm0.01$ \\
 
NGC 985 & $1.87_{-0.03}^{+0.03}$ & $6.79\pm0.09$& $\le0.165$ & $4.24\pm2.60$ & 11.10 & 99.60\% & $18.5_{-6.5}^{+6.5}$ & -$0.02\pm0.01$ \\
  
NGC 2617* & $1.69_{-0.08}^{+0.07}$ & $7.44\pm0.11$ & $\le 0.173$ & $18.45\pm1.03$ & 12.09 & 99.40\% & $35.2_{-10.0}^{+10.5}$ &  $0.07\pm0.01$\\
   
NGC 4051 & $1.97_{-0.01}^{+0.03}$ & $7.10\pm0.09$ & $\le0.11$ & $4.98\pm2.09$ & 12.96 & 99.80\%  & $18.6_{-7.0}^{+6.3}$ & $0.02\pm0.01$ \\
  
NGC 5548 & $1.70_{-0.01}^{+0.01}$ & $7.12\pm0.11$ & $\le 0.211$ & $7.55\pm3.74$ & 7.91 & 95.90\% & $11.7_{-5.3}^{+5.0}$  & $0.02\pm0.01$ \\
   
NGC 5548* & $1.70_{-0.01}^{+0.01}$ & $11.02\pm0.11$ & $\le0.153$ & $14.75\pm9.22$ & 11.60 & 99.20\% & $77.6^{+25.2}_{-24.6}$  & $0.43\pm0.01$ \\
  
NGC 7469* & $1.72_{-0.08}^{+0.05}$ & $10.88\pm0.14$ & $\le 0.162$ & $12.29\pm6.58$ & 11.79 & 99.80\% & $70.0_{-21.4}^{+18.5}$ & $0.42\pm0.01$\\
  
PG 2304+042* & $1.88_{-0.07}^{+0.11}$ & $9.78\pm0.13$ & $\le 0.110 $ & $12.26\pm5.82$ & 11.26 & 99.20\% & $114.5_{-28.8}^{+39.7}$ & $0.33\pm0.01$\\
   \hline
\end{tabular}
\vspace{2mm}
\label{Table 3}
\vspace{-0.3cm}
\tablefoot{*Source with UFO.\\ (1) Source name; (2) rest-frame energy in keV; (3) width in keV; (4) normalization in units of photons cm$^{-2}$s$^{-1}\times10^{-6}$; (5) $\Delta C_{stat}$; (6) significance from MC simulations (see Sect. \ref{Statistical analysis and robustness of results}); (7) EQW in eV; (8) velocity in units of $\rm c$. Uncertainties are at 95\%. EQW uncertainties at 68\%. }
\end{table*}

We adopted MYTorus to model the reprocessed emission from distant circumnuclear material, as it self-consistently accounts for the Compton-scattered continuum and the associated fluorescent Fe K lines. This physically motivated model has become the standard for spectral fitting of local Seyfert galaxies, superseding older infinite-slab approximations such as \texttt{pexrav} \citep{Magdziarz95} or  \texttt{pexmon} \citep{Nandra07} (see \cite{Yaqoob12,Balokovic18} for detailed comparisons). 

Some spectra required additional components to adequately describe the continuum and other emission features. When necessary, the model above was modified, by including the presence of a second power-law to reproduce the soft excess and/or Gaussian emission lines to model the broad component from Fe $K{\alpha}$ with \textsc{kdblur2}, a convolution model applied to the Gaussian emission line to mimic relativistic effects from an accretion disk around the SMBH. 

In some cases, it was also necessary to add Gaussian absorption lines at energies $<6$ keV to obtain a good model for the spectra. These features may represent UFO absorption from other elements
(e.g., Si XIV, SXV, ArXVIII, or Ca XIX) or possible inflows. To distinguish between the two possibilities, a full modeling of the ionized absorbers with photoionization models such as \texttt{Xstar} is needed. However, we defer the analysis of such features to a forthcoming paper, as we focus on detecting UFOs through Fe band features. We illustrate the BM model applied to sources that exhibit a UFO in Appendix \ref{figure modelli}.\\
We verified that none of the X-ray spectra of the radio-loud AGNs in the sample were dominated by the radio-jet contamination; in particular \cite{Madsen15} shows how in 3C~273, the most radio-loud source in the sample, with $R>>100$, the jet contributes significantly in the emission only over $20-30$ keV, which it is outside the analyzed range, therefore it does not affect our modeling.

\subsection{Search for Fe K absorption} \label{sect:abs}

To blindly search for any absorption signature, we added a Gaussian absorption line (zgauss) to the best-fit continuum model, with its energy free to vary in the 7–12 keV rest-frame range, negative normalization, and width also free to vary up to a maximum of 1 keV, to avoid fitting continuum fluctuations. The fit identifies the strongest absorption feature in this band and produces the Best Fit Model (BFM). This procedure follows the same logic as the blind-search method of \citet{Miniutti06}, adopted by \citetalias{Tombesi10} and \citetalias{Matzeu23}. The main difference is that, in those works, the line width is frozen, leaving 2 free parameters, and the $\Delta\chi^2$ or $C_{stat}$ improvement is mapped onto a two-dimensional energy–intensity grid. 

\begin{figure}[h]
    \centering
    \includegraphics[scale=1, angle=0, width=8cm,height=8cm,keepaspectratio]{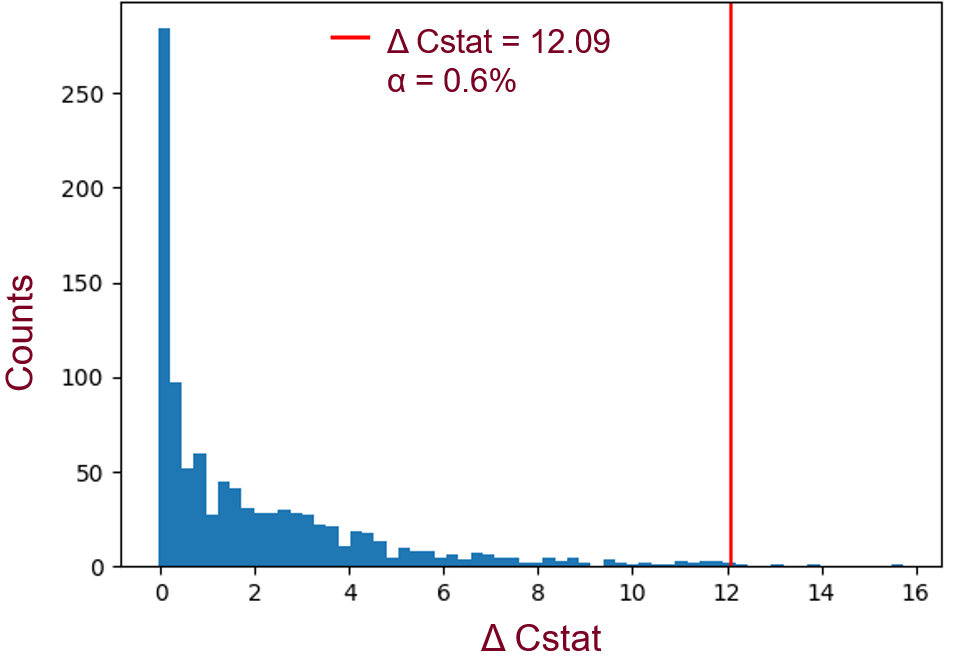}
    \caption{Result of the MC simulations for the spectrum of NGC~2617. The blue histogram shows the distribution of $\Delta C_{stat}$ obtained by fitting the BM and BFM models to the simulated line-less spectra. The red line shows the observed $\Delta C_{stat}$ for this source. $\alpha$ represents the fraction of MC simulations showing a $\Delta C_{stat}$ greater than the observed one. In this case, the absorption feature is detected with a confidence $1-\alpha > 99\%$.}
    \label{cstat}
\end{figure}

In our case, the additional free parameter (line width) makes the construction of such maps impractical; instead, the fit converges to the strongest feature, and its significance is evaluated via the $\Delta C_{stat}$ improvement for 3 degrees of freedom. We adopted $\Delta C_{stat} > 7.815$ as the detection threshold, corresponding to 95\% confidence for 3 degrees of freedom \citep{Lampton76}. For comparison, \citetalias{Tombesi10} and \citetalias{Matzeu23} used $\Delta C_{stat}> 9.21$, corresponding to 99\% confidence for 2 degrees of freedom, since they froze the width parameter. Features exceeding our threshold were then subjected to Monte Carlo (MC) simulations to robustly assess their significance, as described in \Cref{Statistical analysis and robustness of results}. 

In Table \ref{discarded} we list the absorption lines found in the blind search, which show a $\Delta C_{stat}<7.815$ and are therefore considered not significant. Several of these features correspond to UFOs detected in previous works \citep[e.g.,][\citetalias{Tombesi10}]{Gofford13}, but were discarded due to our conservative approach.

\subsubsection{Statistical analysis and robustness of results} \label{Statistical analysis and robustness of results}

Following a robustly conservative approach, in the analysis here presented, we select candidate absorption features only if they meet the statistical criteria for detection with all three background subtraction methods outlined in Sect. \ref{data reduction}, and we report significance results relative to the set up that led to the smallest improvement in the Cash statistic between the model with an absorption line and the one without it. \\
For the features considered significant on the basis of the  $\Delta C_{stat}$ value, we implemented MC simulations to estimate more accurately their statistical significance:

\begin{itemize}
    \item We simulated 1000 EPIC-pn spectra using the \textsc{fakeit} command in XSPEC, assuming the BM and the same exposure times, background levels, and response matrices of the real observation;
    \item We apply and fit both BM and BFM on the fake, line-less spectra, evaluating for every simulation the $\Delta C_{stat}$ between the two models, varying the line energy over the entire 2-12 keV band and looking for the strongest random fluctuations in the simulated line-less spectra;
    \item We compare the values of $\Delta C_{stat}$ of each simulation with the difference in $C_{stat}$ evaluated on the real spectrum of the source and calculate the fraction of simulations where the $\Delta C_{stat}(sim)$>$\Delta C_{stat}(real)$.
\end{itemize}

This procedure aims to estimate the level of significance of a detected absorption line: if more than $\alpha=5\%$ of the 1000 simulations show a Cash statistic improvement greater than the observed $\Delta C_{stat}$, we reject, in a conservative approach, the tentative absorption feature, as it could be a possible noise statistical fluctuation.
Figure \ref{cstat} shows the result of this test applied to the absorption feature detected in NGC~2617, which shows a significance ($1-\alpha$) over 99\%. The significance values associated with the absorption lines detected are shown in Table \ref{Table 3}.

\subsubsection{Uncertainties of the absorption parameters} \label{Uncertainties of the parameters}
In this section, we describe how we estimated the statistical uncertainties on the best-fit parameters of the absorption lines using MC simulations. For each source, we ran a MC simulation that generated 1000 spectra based on the BFM and fit them with it, leaving the absorption line parameters free. We define the confidence interval for each parameter as the interval that encompasses 95\% of the distribution of best-fit values from the 1000 simulations.

We interpreted the absorption lines as blueshifted Fe XXVI, which has a rest-frame energy of 6.97 keV. From the comparison between the observed energy $E_{obs}$, estimated as the median of the distribution of the simulations, and the rest-frame energy of Fe XXVI, we calculated the outflow velocity from this formula, corrected for cosmological redshift:
$\frac{E_{obs}}{E_{FeXXVI}}=\sqrt{(1-\frac{\vout}{\rm c})/(1+\frac{\vout}{\rm c})}$.

Table \ref{Table 3} reports the properties of the absorption lines detected in our sample and the estimation of $\vout$ from the shift in energy. We also compared our results with those obtained using the Bayesian X-ray Analysis tool (BXA; \citep{buchner16}). The comparison of the results from the two methods is shown in Appendix \ref{bxa}. The results on UFO properties are perfectly consistent for all the sources (see Fig. \ref{fig:conf}).

\subsection{Photoionization modeling of Fe K features}
\label{sec:xstar}

In order to characterize the physical properties of the absorbers and confirm the nature of UFOs, we replaced the Gaussian absorption profiles detected with >95\% confidence level with \texttt{Xstar} photoionization tables \citep{BautistaKallman01}, which provide a first-order physical measurements of the absorbing medium of the outflows. 
In particular, we are able to quantify the ionization parameter $\xi$, the column density of the medium $N_H$, and the systemic redshift of the material relative to the observed one, which translates into the outflow velocity $\vout$.

For each source, we used \texttt{Xstar} grids covering turbulent velocities in the range $\sigma_{turb}$ in the span $300-5000$ $\rm km/s$, choosing the table that minimized the residuals of the fitting, since different values of $\sigma_{turb}$ are necessary to properly reproduce the observed widths of the Fe K absorption lines.\\
The best fit values obtained from the fit with \texttt{Xstar} tables are shown in \Cref{Table xstar}. For each source, we first found the value of $v_{\rm turb}$ that best reproduced the detected absorption lines, using a grid that treated $v_{\rm turb} $ as a free parameter; then we selected the tables with the closest fixed values. The \texttt{Xstar} modeling reproduces results comparable with the Gaussian absorption line component of the BFM, identifying the features as blueshifted Fe XXVI absorption line, or a blend of Fe XXV and Fe XXVI. 
For almost all sources, the \texttt{Xstar} modeling led to an improvement in the statistic relative to the Baseline model, with $\Delta C_{stat}>7.815$. In some cases, the \texttt{Xstar} tables fit the residuals better than a Gaussian absorption line model, suggesting that secondary features are modeled together with the main feature. In one case, instead, PG 2304+042, the improvement in the fit with the \texttt{Xstar} table is $\Delta C_{stat}<7.815$. This source, as shown in Fig. \ref{ngc7469_pg2304}, exhibits a very deep, narrow absorption feature, with EQW>100 eV, and even the grids with the smallest $v_{turb}$ (i.e., 300 $\rm km/s$) struggle to fit the observed residuals entirely.
The narrow, deep profile in PG~2304+042 may require a multi-zone absorber or a higher-resolution grid to be fully reproduced; nonetheless, the Gaussian parameters are fully consistent with blueshifted Fe XXV/XXVI at $\vout\sim0.33\rm c$.

\section{Results and discussion}
\label{Sec. 5}
\begin{figure*}[t]
\centering
    \includegraphics[scale=1, angle=0, width=8.5cm,keepaspectratio]{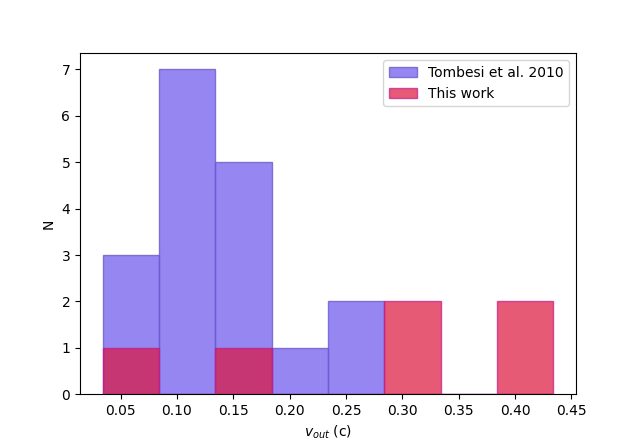}
    \hspace{0.7cm}
    \includegraphics[scale=1, angle=0, width=8.5cm,keepaspectratio]{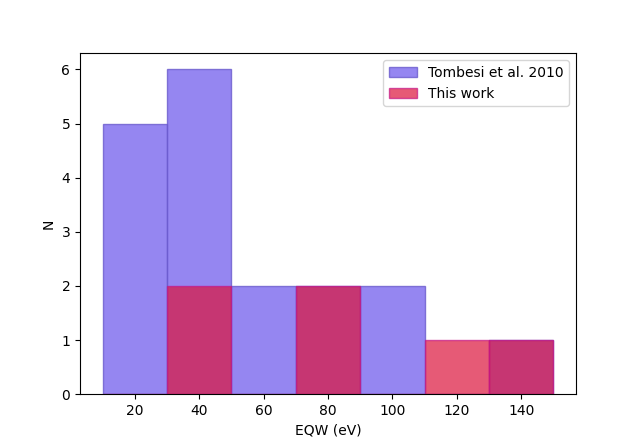}
    \caption{Left: Comparison between the UFO velocity distributions of our sample (magenta) with the \citetalias{Tombesi10} sample (purple). Right: Same for the EQW distribution.}
    \label{vel_eqw}
\end{figure*}

\subsection{Extending the high end of the UFO velocity distribution}
Table \ref{Table 3} presents the best-fit parameter values of the analyzed absorption lines, along with their corresponding velocities and uncertainties, at 95\% significance level. Features classified as UFO based on their velocity are marked with * symbol.
Although we detected absorption lines with high significance in 8 out of 33 sources, we classified as UFO only absorption features blueshifted by $\vout>10000 ~\kms$, as in the work of \citetalias{Tombesi10}, so NGC~985 (which has an inflow), NGC~4051, and one of the lines in NGC~5548 are excluded. 
In the following, we compare our work with that of \citetalias{Tombesi10}, as both utilize the highest-quality spectra available in the \xmm archive. Both samples are predominantly composed of nearby Seyferts, despite our sample spanning a larger luminosity and redshift range. 

We note that comprehensive analyses of UFO properties across larger, heterogeneous compilations have been presented in recent literature. \citet{Gianolli2024} (\citetalias{Gianolli2024} hereafter), combining three independent subsamples, performed a comprehensive population study of UFO properties across a wide range of redshift and luminosity; we compare our results with their correlations in \Cref{sec:physical_prop}. \citet{Yamada2024} assembled a legacy database of 583 X-ray winds in 132 AGN at $z=0-4$, classifying them into UFOs, low-ionization UFOs, and warm absorbers. \citep{Laurenti26} studied 122 robust UFO detections, exploring for the first time the dependence of UFO properties on AGN classes. 
In this work, instead, we perform a homogeneous blind search in a previously unexplored energy range ($10-12$ keV) within a single flux-limited sample.

We report the presence of UFOs in six AGN corresponding to a fraction of $f_{UFO}=(18.2^{+11.0}_{-7.3})\%$, where the uncertainties are computed with the Gehrels formulae \citep{Gehrels1986} for low number statistics. We note that analyzing a single observation per source, while avoiding source-dependent biases, reduces the probability of catching a transient UFO in any given target. To quantify this effect, we compare with the strategy of \citetalias{Tombesi10}, which analyzed an average of $\sim2.4$ observations per source. If we restrict the \citetalias{Tombesi10} sample to UFOs detected only in the longest observation of each source — mirroring our selection — the detection rate drops from $\sim35$\% to $\sim14$\% (6 UFOs over 42 sources), fully consistent with our measured fraction. 
Rescaling by the average number of observations recovers the original \citetalias{Tombesi10} rate ($14\%\times2.4\sim34\%$), supporting the interpretation that the per-observation detection probability is comparable between the two studies, and that the difference in the observed UFO fraction is driven by the number of independent epochs sampled rather than by differences in sensitivity or sample properties. This also underscores the impact of UFO variability and duty cycle on single-epoch detection fractions, and motivates the multi-epoch extension of this work discussed in Sect. \ref{conclusions}.

Figure \ref{vel_eqw} (left) shows the comparison between the velocity distribution obtained from our sample and the one found by \citetalias{Tombesi10}. It is evident that our distribution is shifted toward higher velocities, thanks to the extension of the UFO search up to 12~keV. A KS test between the two distributions yields a p-value of 0.028, corresponding to a confidence level of $97.2\%$ that they are statistically different. It is important to note that we would not have been able to detect such features if our analysis had been limited to 10~keV, as done in \citetalias{Tombesi10} and \cite{Gofford13}. In fact, Mkn~766, NGC~5548, NGC~7469, and PG~2304+042 show UFOs with velocities higher than $0.3\rm c$, which are absorption features detected above 10~keV. 

As previously discussed, revealing UFOs with $\vout>0.3\rm  c$ (i.e., $\vout>90000$ km/s), is difficult as the absorption features are highly blueshifted and, therefore, are located at energies where the effective area of the operational X-ray telescopes decreases. 
Furthermore, it is well known that the EPIC-pn camera calibration is less constrained in this high-energy regime. To validate the robustness of our detections above 10 keV against potential instrumental artifacts, we cross-checked our results using simultaneous \nustar observations, which offer good sensitivity and calibration in the hard X-ray band, albeit with lower spectral resolution. The results of this validation are detailed in \autoref{app:Nustar}.

Given the difficulty, we can study these highly blueshifted features only with an accurate characterization of the background, and if the signal-to-noise ratio of the observation is very high. Remarkably, \cite{Luminari23} presented the discovery of an extreme UFO at $E\sim11~\rm  keV$ with $3\sigma$ confidence level in the nearby Seyfert galaxy NGC 2992 caught during a high-flux X-ray state.
It is worth noting that winds with $\vout>0.3 \rm c$ have been observed in high-z QSOs (see \cite{Chartas02,Chartas21}, \citetalias{Chartas21} hereafter), since UFOs absorption features fall, due to the redshift of the sources, into the observability range of \xmm and Chandra.

We detected a UFO in NGC~5548 and NGC~7469 for the first time. Both sources reveal a highly complex, cold, and low-ionized absorption in the soft band (\cite{Kaastra14}, \cite{Behar2017}), and the presence of a UFO lends support to a multiphase nuclear outflow scenario. Moreover, NGC~5548 exhibits two absorption lines, indicating that the detected UFOs are likely composed of different components at various velocities, thereby validating the hypothesis that AGN outflows are inherently multiphase.

\begin{figure}[ht]
    \centering
    \includegraphics[scale=1, angle=0, width=9cm,keepaspectratio]{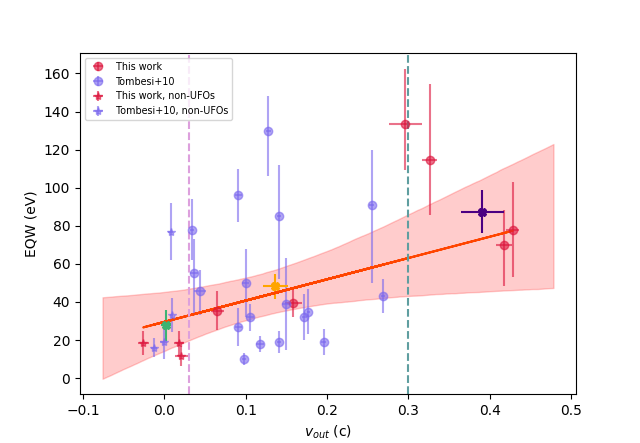}
    \caption{Comparison between the UFOs' equivalent width and outflow velocity distributions of our sample (in red) with the sample from \citetalias{Tombesi10} (in blue). The dashed plum line marks the threshold of 0.03c, while the dashed turquoise line marks 0.3c. The green, orange, and purple points are the mean of the velocity and EQW in the three velocity bins.
    {The red line and shaded area show the linear best fit and its 95\% confidence interval computed with LINMIX, taking into account errors on both $\vout$ and EQW.}}
    \label{eqw vs v}
\end{figure}

\begin{figure}[ht]
    \centering
    \includegraphics[scale=1, angle=0, width=8.25cm,keepaspectratio]{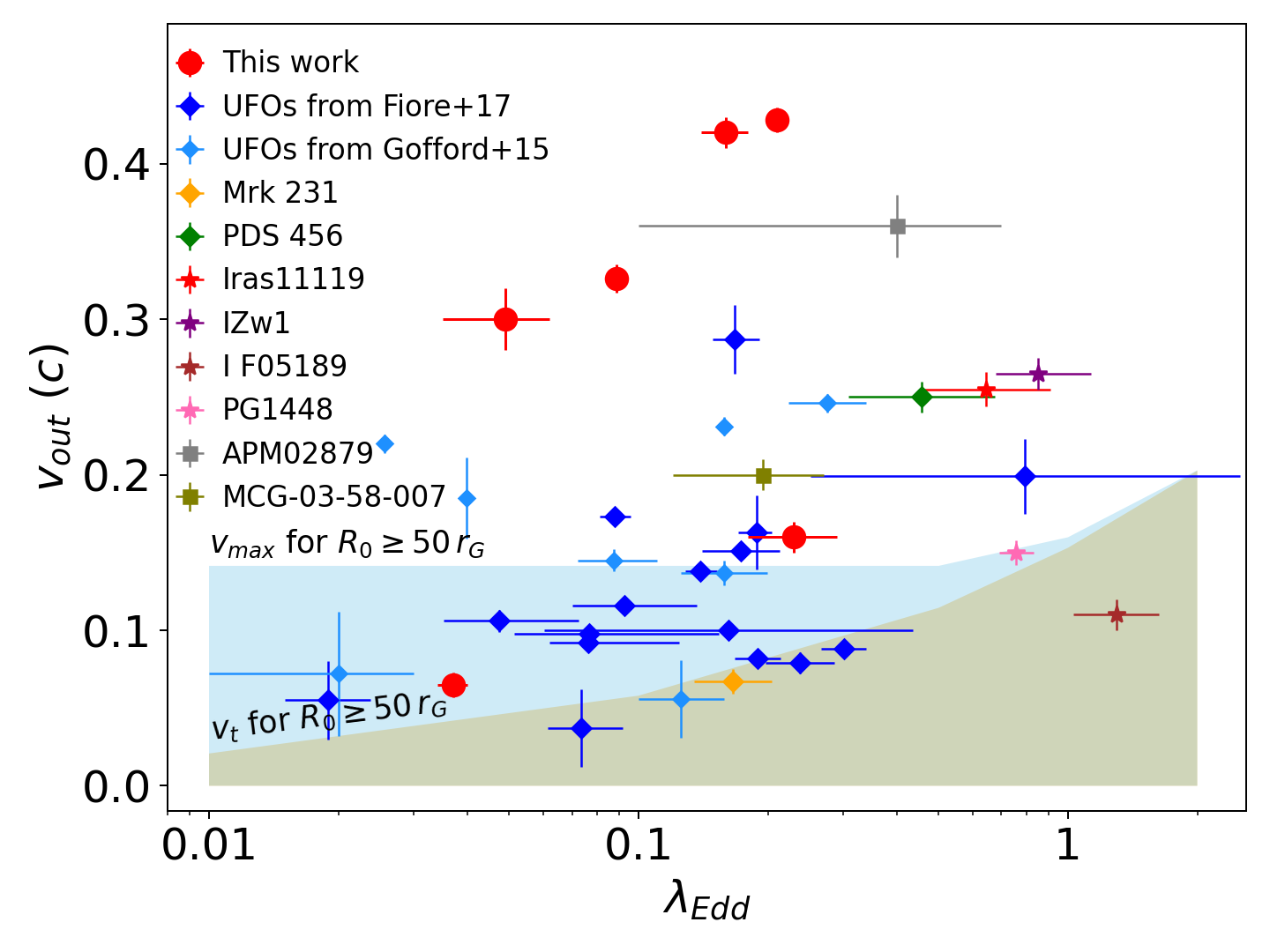}
    \caption{Comparison between UFO velocities from the literature (see label) and the results of the present work (red circles) as a function of $\lambda_{Edd}$ (from \cite{Luminari21}). The theoretical limit, according to the terminal (maximum) velocity $v_t$ ($v_{\rm max}$), for $R_0 \geq 50~\rm r_G$ is shown in the green (light-blue) shaded area.}
    \label{lambda vs v}
\end{figure}

Figure \ref{vel_eqw} (right) presents the distribution of equivalent widths (EQW) for UFOs in our sample compared with \citetalias{Tombesi10}. 
Even if the difference between the two samples is not statistically significant (the KS test yields a significance of only 70\%), the EQWs obtained in this work are larger than 40-50 eV, while the \citetalias{Tombesi10} distribution goes down to 10-20 eV, and $\sim40$ eV is close to the median EQW. 
The higher EQWs found in this work can possibly be explained by Fig. \ref{eqw vs v}. Here we show the EQWs of the absorption lines versus the velocity of the nuclear winds of both samples. 
We divided the plane into three bins, with limits of $\vout < 0.03\rm c$ and $\vout > 0.3\rm c$, separating the low-velocity UFOs, those with velocities in the typical range, and those detected at high velocity via the 12 keV analysis extension. We calculated the global mean value of $\vout$ and EQWs for UFOs in each bin and tested the linear correlation between the two quantities, first with the Pearson test, which reported a confidence level over 99\%, and then with LINMIX \citep{Kelly07}, a Bayesian tool for linear regression that takes into account errors on both the x and y axes. The red line represents the linear best fit, while the shaded area represents its 95\% confidence interval. The latter shows a weak positive correlation between outflow velocity and EQW. Having 2/3 of the detected UFOs in the high-velocity regime means that we are also shifting the EQW distribution toward higher values. Indeed, Table \ref{Table 3} shows that we are sensitive to weak features with EQW$\sim10-20~\rm eV$, but these are all in the low velocity regime.

\begin{table*}[t]
\centering
\caption{Photoionization modeling of the detected UFO features with \texttt{Xstar} tables.}
  \renewcommand{\arraystretch}{1.25}
\begin{tabular}{c c c c c c}
\hline
Source  & $\sigma_{turb}$ & $log(\NH)$ & $log(\xi)$ & $\vout$ & $\Delta$ Cstat\\
\hline
(1) & (2) & (3) & (4) & (5) & (6)\\
\hline
\hline

Mkn 766 & 3000 & $22.52_{-0.51}^{+0.70}$ & $4.22_{-0.58}^{+0.47}$ & $0.32_{-0.01}^{+0.01}$ & 14.59 \\
   
Mrk 110 & 3000 & $>22.42$ & $>3.38$ & $0.18_{-0.03}^{+0.02}$ & 8.68 \\
  
NGC 2617 & 1000 &$ >23.67$ & $>4.78$ & $0.07_{-0.01}^{+0.01}$ & 9.45 \\

NGC 5548 & 5000 & $22.16_{-0.28}^{+0.42}$ & $3.80_{-0.20}^{+0.44}$ & $0.47_{-0.01}^{+0.01}$ & 13.27 \\
  
NGC 7469 & 3000 & $22.15_{-0.24}^{+0.08}$ & $3.81\pm0.42$ & $0.45_{-0.01}^{+0.01}$ & 9.81 \\
  
PG 2304+042 & 1000 &$ >23.30$ & $>2.92$ & $0.33_{-0.01}^{+0.01}$ & 4.35 \\
   \hline
\end{tabular}
\vspace{2mm}
\label{Table xstar}
\vspace{-0.3cm}
\tablefoot{(1) Source name; (2) intrinsic velocity broadening of the \texttt{Xstar} grid in $\rm km~ \rm s^{-1}$; (3) column density in $\rm cm^{-2}$; (4) gas ionization state both in log scale in $\rm erg$ $\rm cm$ $\rm s^{-1}$; (5) corresponding outflow velocity in v/c; (6) $\Delta C_{stat}$.}
\end{table*}

Figure \ref{lambda vs v}, adapted from \cite{Luminari21}, shows the outflow velocity of UFOs versus $\lambda_{Edd}$ of the AGN for our sample (in black) and compared to \cite{Fiore17}, \cite{Gofford15} and several individual sources (see Appendix E of \cite{Luminari21} for the complete list of references). While it appears that the $\vout$ increases with the accretion rate of the AGNs, our sources appear to have faster UFOs than local sources with higher $\lambda_{Edd}$. The figure also shows the allowed velocity ranges in a radiative driving framework, including special relativistic effects \citep{Luminari20}, for the terminal velocity $v_t$ and the maximum velocity $v_{\rm max}$. 
The latter represents an upper limit for the observable velocity, for a nuclear wind launched at $r\geq50~\rm R_G$, where $R_G$ is the gravitational radius, assumed in \cite{Luminari21} as the lower bound for the UFO launching radius. The high velocities of the UFOs in our sample can hardly be explained uniquely with a radiative driving model, suggesting the involvement of other launching mechanisms, such as MHD (\cite{Laha21}).

Most of the sources in our sample were previously analyzed by other works in search of UFOs. However, the six UFOs identified in our analysis were never detected before (but see below). In Table \ref{Table 6.1}, we present a comparison of our results with those published in the literature for the same sources. In some cases (5 out of 16), our results differ from those published on the same observations, either because we identify absorption features that were not originally discovered or vice versa. 

\begin{figure*}[t]
    \centering \includegraphics[width=\textwidth,keepaspectratio]{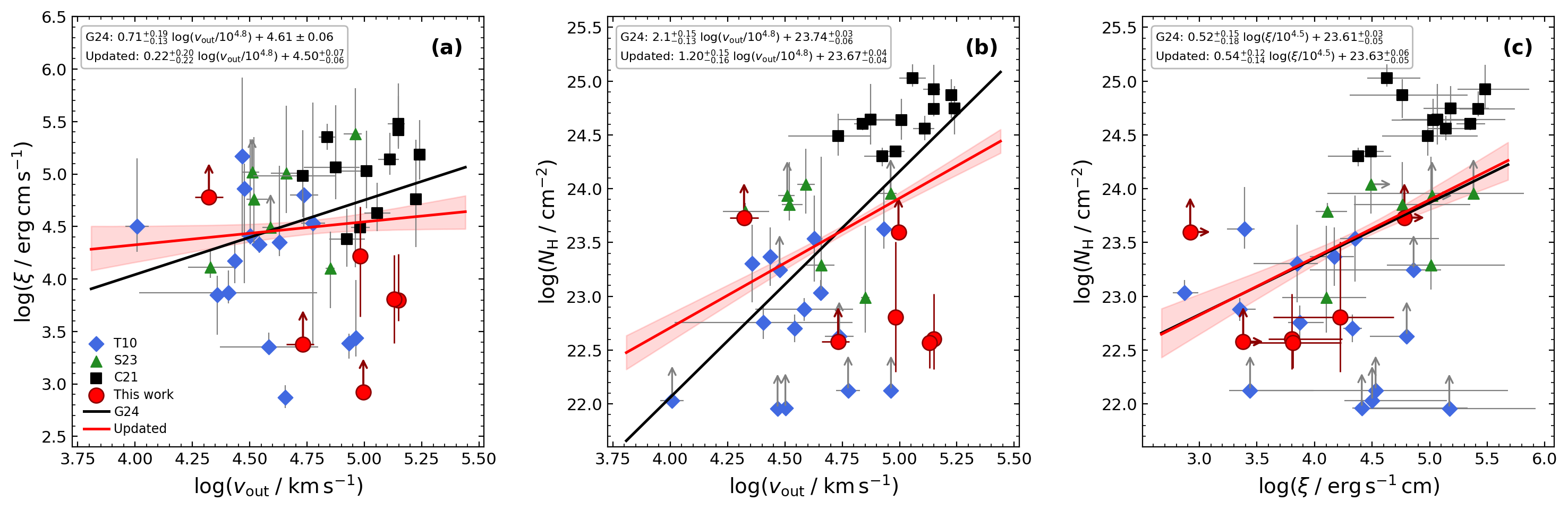}
    \caption{Correlations between observed UFO parameters: ionization parameter $log(\rm \xi)$ vs. outflow velocity $log(\vout)$ (panel a), column density $log(\NH)$ vs. $log(\vout)$ (panel b), and $log(\NH)$ vs. $log(\rm \xi)$ (panel c). Blue diamonds, green triangles, and black squares represent the \citetalias{Tombesi10}, \citetalias{Matzeu23}, and \citetalias{Chartas21} samples from \citetalias{Gianolli2024}, respectively; red circles are the detections from this work. The black solid lines show the original best-fit relations, while the red solid lines and shaded areas show the updated fits and their 68\% confidence intervals after including our new detections.}
    \label{fig:gian24}
\end{figure*}

There are several possible explanations for this discrepancy. (1) we performed the fit on a larger energy range, enabling an extension of the velocity parameter space under study and, in some cases, a better determination of the underlying continuum; (2) we have investigated the detection and properties of the absorption features against all different available methods to determine the instrument particle background, which constitute the dominant background EPIC-pn component in hard X-rays; (3) we have left the line width as a free parameter and used an initial threshold of 95\% confidence for three parameters ($\Delta C_{stat}>7.815$), while the typical approach is to adopt a narrow, unresolved line, and (4) finally, we are using state-of-the art spectral analysis methods, which have been perfected in the last few years such as the proper statistical treatment of the Poissonian maximum likelihood \citep{Kaastra_2017}, optimal binning \citep{KaastraBleeker16}, Monte Carlo and Bayesian methods to estimate the posterior absorption feature parameter distributions (\cite{Buchner14}). 

An interesting example is Mkn~766. Several low-velocity features 
have been reported in the past for this source, based on observations different from the one analyzed here ($\vout=0.044\pm0.007\rm c$, $0.091\pm0.004\rm c$ in \citetalias{Tombesi10} and $\vout=0.061\pm0.008\rm c$, $0.017\pm0.004\rm c$ in \citealt{Gofford13}). On the contrary, in the same observation analyzed here, a significant feature was identified at the same energies ($\sim9.5$ keV) both in \citetalias{Tombesi10} and in \citealt{Pounds03mrk766} but interpreted as a $9.28$ keV H-like Iron edge outflowing at $\sim0.05\rm c$ and linked to a transient X-ray flare, rather than a high-velocity UFO. This highlights the importance of an accurate continuum modeling above the absorption feature, which allows us to distinguish between the two scenarios.\\
It is worth noting also that in some of the sources (e.g., Mrk 509) in which previous studies found UFOs, we actually detect UFO-related absorption lines, but with insufficient significance (see Sect. \ref{Statistical analysis and robustness of results}). They are shown in Table \ref{discarded} in the Appendix.
\\
Finally, by examining the literature, it is possible to find cases in which different results are obtained when comparing observations of the same source taken at different epochs. For example, 3C~120 has a UFO with $\vout=0.161\rm c$ \citep{Tombesi14}, and NGC 4051 has a faster UFO with $\vout=0.150\rm c$ \citepalias{Tombesi10}, both detected in observations not analyzed in this work. This is expected and highlights the transient nature of UFOs.

\subsection{Physical properties of the UFOs}
\label{sec:physical_prop}

Using the results of the fit with \texttt{Xstar} photoionization tables, we can place the physical properties of the UFOs detected in this work in the broader context of AGN wind studies, by updating the correlations between the observed outflow parameters ($\xi$, $\NH$ and $\vout$) presented by \citetalias{Gianolli2024}, who combined the \citetalias{Tombesi10} sample with results from SUBWAYS \citepalias{Matzeu23}, and the high-z sample from \citetalias{Chartas21}. \autoref{fig:gian24} shows the three correlation planes from their Fig. 4, now including our new detections (red circles). The best-fit relations reported by \citetalias{Gianolli2024} are shown as black lines, while the updated fits incorporating our data are shown in red, with the shaded area representing the 68\% confidence interval. 

The most striking change is in panel (a): the log($\xi$)– log($\vout$) correlation flattens significantly with slope $0.22^{+0.20}_{-0.22}$ (with respect to $0.71^{+0.19}_{-0.13}$) and is no longer statistically significant. This is driven by the fact that several of our UFOs have high velocities ($\vout > 0.3\rm c$) but moderate ionization parameters ($log(\xi)\sim3.8-4.2$), populating a previously unexplored region of the parameter space. 
This suggests that the outflow velocity is not primarily governed by the gas ionization state, as would be expected in radiative-driving scenarios, lending further support to MHD-driven acceleration models \citep[e.g.,][]{Fukumura17,Fukumura22}.

In panel (b), the $log(\NH)-log(\vout)$ relation remains positive but becomes slightly flatter (slope $1.20^{+0.19}_{-0.15}$ versus $2.1^{+0.15}_{-0.13}$), consistent with the fact that our high-velocity detections, enabled by the extension to 12 keV, do not require the very large column densities ($log(\NH/\rm cm^{-2}) > 24$) typical of the \citetalias{Chartas21} high-redshift quasars. The $log(\NH)-log(\xi)$ correlation in panel (c) remains essentially unchanged (slope $0.54^{+0.15}_{-0.13}$ versus $0.52^{+0.15}_{-0.18}$), as our sources populate the same ranges as the \citetalias{Tombesi10} and \citetalias{Matzeu23} samples in this plane. 

It is worth noting that the original \citetalias{Gianolli2024} correlations in the $log(\xi)- log(\vout)$ and $log(\NH)-log(\vout)$ planes are largely driven by the high-redshift quasars sample, whose absorber parameters are subject to both significant systematic uncertainties, being strongly model-dependent, and observationally selection biased, as the lower spectral quality of high-z data restricts detections to the strongest absorption features, favoring intrinsically high $\NH$ and $\xi$ values. The flattening we observe when our local, high-quality detections are included highlights the need for a homogeneous reanalysis of the UFO population across redshift.

Finally, the outflow velocity measured by Gaussian fitting is comparable with the findings of \texttt{Xstar} tables, confirming that the observed features are due to either Fe XXV or XXVI $K\alpha$ lines, or a blend of the two. A Kolmogorov-Smirnov test applied to the velocity results shows that the photoionization and the Gaussian modeling of the Fe K absorption features have the same distribution, with a 96.3\% confidence level. 

\subsection{UFO versus AGN properties} \label{UFO vs. AGN accretion properties}

\begin{figure*}[t]
    \centering
    \includegraphics[scale=1, angle=0, width=7cm,keepaspectratio]{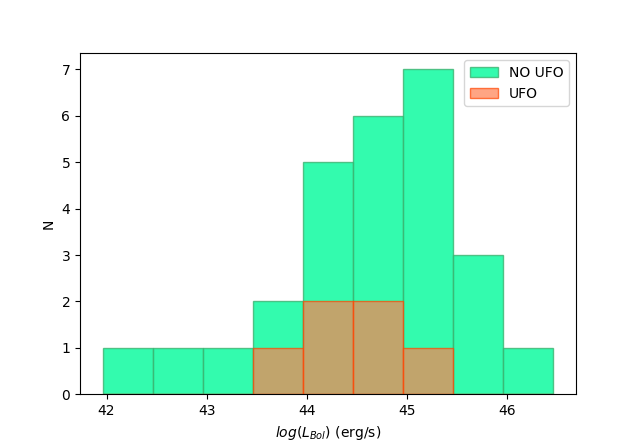}
    \hspace{0.3cm}
    \includegraphics[scale=1, angle=0, width=7cm,keepaspectratio]{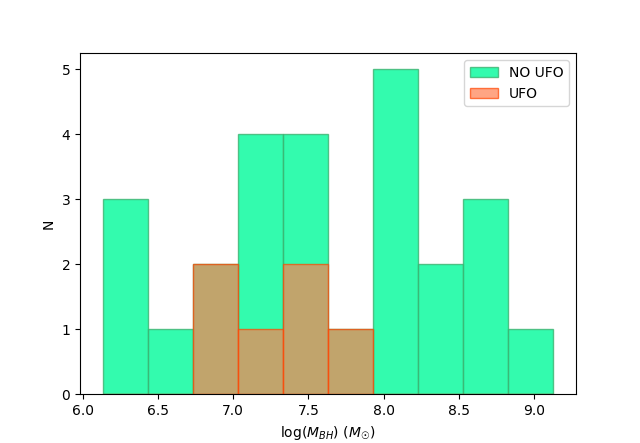}
    \vspace{0.3cm}
    \includegraphics[scale=1, angle=0, width=7cm,keepaspectratio]{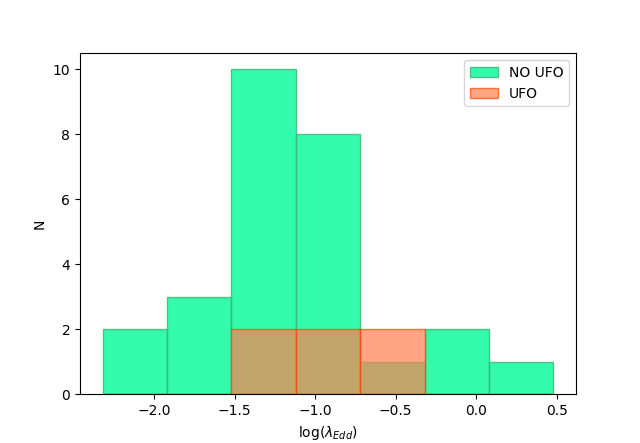}
    \hspace{0.3cm}
    \includegraphics[scale=1, angle=0, width=7cm,keepaspectratio]{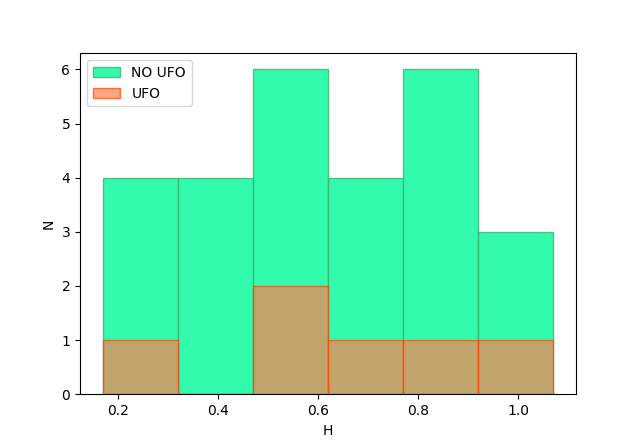}
    \caption{Distribution of AGN with (orange) and without UFOs (green) with respect to their bolometric luminosity, BH mass, Eddington ratio, and Hhardnessparameter (see Table \ref{Table 4.1}).}
    \label{L BOL}
\end{figure*}

One of the goals of our analysis consist to understand if the incidence of UFOs depends on the AGN properties, such as the mass of the SMBHs, as $L_{\rm bol}$, $\lambda_{\rm Edd}$, and H. We divided our sample into two groups: one with only sources showing UFO detections and the other with sources that did not. Then we studied the distributions of the two groups with respect to accretion properties (as shown in Fig. \ref{L BOL}) of AGN and verified whether there are significant differences between the two groups using a KS test.

We emphasize that such a comparison between UFO-hosting and non-UFO-hosting sources is restricted to the 33 AGN in our parent sample, for which accretion properties ($L_{\rm bol}$, H, $\lambda_{\rm Edd}$) were derived self-consistently from simultaneous \xmm OM and EPIC data by \citet{Svoboda17}. Our sample is not designed to provide a comprehensive census of the UFO population, and our results on UFO incidence versus AGN properties should not be extrapolated to the general AGN population. For broader analyses of UFO–accretion correlations across heterogeneous compilations, we refer the reader to \citetalias{Gianolli2024}, \citet{Yamada2024} and \citet{Laurenti26}, for example. Within this limited scope, we investigate whether the incidence of UFOs in our sample depends on the AGN properties.

The KS lends to the following confidence levels: 1-p=73\% for $L_{\rm bol}$, 1-p=87\% for the mass of the SMBHs, 1-p=30\% for the $\lambda_{\rm Edd}$, and 1-p=4\% for H values. Therefore, in none of the cases can we state that the two samples come from different distributions, as none of the values exceeds the 95\% confidence level. It is also important to highlight the small size of the UFO sample in this work, making it challenging to derive firm statistical conclusions.

These results conflict with what has been found for stellar BHBs: our study, in fact, indicates that the launch of UFOs in AGNs requires physical conditions of the accretion disk and corona that differ from those underpinning the
production of outflows in stellar black holes, indicating a dependence more on the duty cycle of the sources than on their accretion properties.\\ Although our analysis found no UFOs in AGNs with $M_{BH}>10^8$, the work of \citetalias{Matzeu23} revealed several winds in this mass range, suggesting this is just a small sample size effect and there is no real dependence between this AGN property and the presence of UFOs. 

\section{Conclusions} 
\label{conclusions}
We have performed an investigation on the presence and properties of UFOs, the most powerful, highly ionized nuclear AGN winds, in a sample of bright local Seyfert galaxies.
We collected high-quality \xmm pn spectra for 33 sources. The sources (reported in Table \ref{Table 4.1}) were selected from the same catalog studied by \cite{Svoboda17}, choosing a threshold in flux of $F_{2-12 \rm  keV}=1.5\times10^{-11}$ erg s$^{-1}$ cm$^{-2}$.
We searched for absorption features between $7-12$ keV rest-frame, characteristic of UFOs, by adding Gaussian absorption lines to the spectral continuum models. Furthermore, we implemented MC simulations to evaluate the statistical significance of these absorptions, as well as the confidence levels for the energy, width, and strength of each line. The main results of this study are summarized below:

\begin{enumerate}
    \item We report highly significant (95\% confidence level) detections of UFOs in six sources of the sample, corresponding to a fraction of $f_{UFO}=(18.2^{+11.0}_{-7.3})\%$. We highlight that these winds were previously undetected. 
    \item The extension to $E>10$ keV in the search for UFO-related absorption features has led to the detection of four UFOs with very high velocities ($\vout>0.3\rm c$) in Mkn~766, NGC~5548, NGC~7469, and PG~2304+042.
    The reliability of the highest-velocity detection methodology was successfully validated against simultaneous \nustar data for NGC~7469 (\Cref{app:Nustar}).
    \item The photoionization modeling of Fe K features through \texttt{Xstar} tables agrees with the results obtained with the Gaussian modeling, confirming our results.
    \item The common presence of UFOs of such high velocity indicates that several distinct acceleration mechanisms may be operating in AGNs. This also suggests a broader velocity distribution for X-ray winds, which may affect the role of UFOs in AGN feedback processes.
    \item The comparison of our results with the outcomes of previous studies on different observations allowed us to monitor the intrinsic variability UFOs in AGNs over time, confirming that the presence and velocity of these highly ionized nuclear winds change on long timescales. 
    \item The comparison of the physical properties of our UFO detections with the population study of \citetalias{Gianolli2024} shows that the addition of high-velocity, moderate-ionization UFOs from our sample significantly flattens the $log(\xi)-log(\vout)$ correlation. The $log(\NH)-log(\vout)$ relation is also moderately flatter, while the $log(\NH)-log(\xi)$ correlation remains unchanged.
    \item We investigated the possible correlations between the presence and characteristics of the detected UFOs and AGN properties. Our analysis shows no particular distinctions between AGNs that exhibit UFOs and those that do not, in our sample, on the basis of the mass of SMBHs, $L_{Bol}$, $\lambda_{EDD}$, and H, in contrast with what was found for X-ray nuclear winds in stellar BHs. This could suggest that all AGNs may host UFOs detectable only in specific epochs. Their identification could also depend on the circumnuclear environment or the inclination with respect to the line of sight, which we did not investigate in this work. 
\end{enumerate}

In summary, this study represents the first exploratory, systematic search for very high-velocity ($\vout> 0.3 \rm c$) UFOs in the $10-12$ keV band of local AGNs, using the deepest available CCD-resolution spectra from the \xmm archive. Our results demonstrate that such extreme velocity components are present in a non-negligible fraction of local bright AGNs, motivating dedicated follow-up at higher spectral resolution. The XRISM Resolve micro-calorimeter, with its well-calibrated response up to $\sim11.5$ keV and spectral resolution superior to EPIC-pn by more than an order of magnitude, is ideally suited to confirm and characterize these features. The XRISM archive is rapidly growing with long observations of several bright Seyferts in our sample, and a systematic search for $>10$ keV UFOs using Resolve data is an ongoing project that will be presented in a forthcoming paper.

A complementary approach is to target high-redshift QSOs, where the cosmological redshift brings the $>10$ keV absorption features into the standard observing band of current X-ray telescopes. This strategy is being pursued by the WISSHFUL XMM-Newton Heritage Programme\footnote{\url{https://sites.google.com/inaf.it/wisshfulproject}} \citep[][Borrelli et al. in preparation]{Lanzuisi26sub}, which is obtaining very deep observations of luminous QSOs at $z = 2-4$, at both BH growth and star-formation peaks, providing an independent route to systematically characterize the fastest components of AGN winds at Cosmic Noon.

Moreover, we plan to extend this work by analyzing multiple high-quality observations of bright AGNs available in the \xmm archive, systematically extending the UFO search up to 12 keV to identify the most powerful UFOs and study their variability and duty cycle over time.

\begin{acknowledgements} 
LB and GL acknowledge support from the Large Program ``DELUX'' of the “Ricerca Fondamentale 2024” INAF program.
EP  acknowledges funding from the European Union – Next Generation EU, PRIN/MUR 2022 (2022K9N5B4).
JS acknowledges GACR support from the project 26-226145.

\end{acknowledgements}

\bibliographystyle{aa}
\bibliography{references}

@string{aap = {A\&A}}

@string{aj = {AJ}}

@string{apj = {ApJ}}

@string{apjl = {ApJL}}

@string{apjs = {ApJS}}

@string{araa = {Annu. Rev. Astron. Astrophys.}}

@string{mnras = {MNRAS}}

@string{pasj = {PASJ}}

@string{pasp = {PASP}}

@ARTICLE{Buchner14,
       author = {{Buchner}, J. and {Georgakakis}, A. and {Nandra}, K. and {Hsu}, L. and {Rangel}, C. and {Brightman}, M. and {Merloni}, A. and {Salvato}, M. and {Donley}, J. and {Kocevski}, D.},
        title = "{X-ray spectral modelling of the AGN obscuring region in the CDFS: Bayesian model selection and catalogue}",
      journal = {\aap},
         year = 2014,
        month = apr,
       volume = {564},
          eid = {A125},
        pages = {A125},
          doi = {10.1051/0004-6361/201322971},
archivePrefix = {arXiv},
       eprint = {1402.0004},
 primaryClass = {astro-ph.HE},
       adsurl = {https://ui.adsabs.harvard.edu/abs/2014A&A...564A.125B}
}

@ARTICLE{BautistaKallman01,
   author = {{Bautista}, M.~A. and {Kallman}, T.~R.},
    title = "{The XSTAR Atomic Database}",
  journal = {\apjs},
     year = 2001,
    month = may,
   volume = 134,
    pages = {139-149},
      doi = {10.1086/320363},
   adsurl = {http://adsabs.harvard.edu/abs/2001ApJS..134..139B}
}

@ARTICLE{Tombesi13,
   author = {{Tombesi}, F. and {Cappi}, M. and {Reeves}, J.~N. and {Nemmen}, R.~S. and 
	{Braito}, V. and {Gaspari}, M. and {Reynolds}, C.~S.},
    title = "{Unification of X-ray winds in Seyfert galaxies: from ultra-fast outflows to warm absorbers}",
  journal = {\mnras},
archivePrefix = "arXiv",
   eprint = {1212.4851},
 primaryClass = "astro-ph.HE",
     year = 2013,
    month = apr,
   volume = 430,
    pages = {1102-1117},
      doi = {10.1093/mnras/sts692},
   adsurl = {http://adsabs.harvard.edu/abs/2013MNRAS.430.1102T}
}

@ARTICLE{Tombesi10,
   author = {{Tombesi}, F. and {Cappi}, M. and {Reeves}, J.~N. and {Palumbo}, G.~G.~C. and 
	{Yaqoob}, T. and {Braito}, V. and {Dadina}, M.},
    title = "{Evidence for ultra-fast outflows in radio-quiet AGNs. I. Detection and statistical incidence of Fe K-shell absorption lines}",
  journal = {\aap},
archivePrefix = "arXiv",
   eprint = {1006.2858},
 primaryClass = "astro-ph.HE",
     year = 2010,
    month = oct,
   volume = 521,
      eid = {A57},
    pages = {A57},
      doi = {10.1051/0004-6361/200913440},
   adsurl = {http://adsabs.harvard.edu/abs/2010A%26A...521A..57T}
}

@ARTICLE{Gofford13,
   author = {{Gofford}, J. and {Reeves}, J.~N. and {Tombesi}, F. and {Braito}, V. and 
	{Turner}, T.~J. and {Miller}, L. and {Cappi}, M.},
    title = "{The Suzaku view of highly ionized outflows in AGN - I. Statistical detection and global absorber properties}",
  journal = {\mnras},
archivePrefix = "arXiv",
   eprint = {1211.5810},
 primaryClass = "astro-ph.HE",
     year = 2013,
    month = mar,
   volume = 430,
    pages = {60-80},
      doi = {10.1093/mnras/sts481},
   adsurl = {http://adsabs.harvard.edu/abs/2013MNRAS.430...60G}
}

@ARTICLE{Pounds03mrk766,
       author = {{Pounds}, K.~A. and {Reeves}, J.~N. and {Page}, K.~L. and {Wynn}, G.~A. and {O'Brien}, P.~T.},
        title = "{Fe K emission and absorption features in XMM-Newton spectra of Markarian 766: evidence for reprocessing in flare ejecta}",
      journal = {\mnras},
         year = 2003,
        month = jul,
       volume = {342},
       number = {4},
        pages = {1147-1155},
          doi = {10.1046/j.1365-8711.2003.06611.x},
archivePrefix = {arXiv},
       eprint = {astro-ph/0302151},
 primaryClass = {astro-ph},
       adsurl = {https://ui.adsabs.harvard.edu/abs/2003MNRAS.342.1147P}
}

@ARTICLE{Ferrarese00,
   author = {{Ferrarese}, L. and {Merritt}, D.},
    title = "{A Fundamental Relation between Supermassive Black Holes and Their Host Galaxies}",
  journal = {\apjl},
   eprint = {astro-ph/0006053},
     year = 2000,
    month = aug,
   volume = 539,
    pages = {L9-L12},
      doi = {10.1086/312838},
   adsurl = {http://adsabs.harvard.edu/abs/2000ApJ...539L...9F}
}

@ARTICLE{Gebhardt00,
   author = {{Gebhardt}, K. and {Bender}, R. and {Bower}, G. and {Dressler}, A. and 
	{Faber}, S.~M. and {Filippenko}, A.~V. and {Green}, R. and {Grillmair}, C. and 
	{Ho}, L.~C. and {Kormendy}, J. and {Lauer}, T.~R. and {Magorrian}, J. and 
	{Pinkney}, J. and {Richstone}, D. and {Tremaine}, S.},
    title = "{A Relationship between Nuclear Black Hole Mass and Galaxy Velocity Dispersion}",
  journal = {\apjl},
   eprint = {astro-ph/0006289},
     year = 2000,
    month = aug,
   volume = 539,
    pages = {L13-L16},
      doi = {10.1086/312840},
   adsurl = {http://adsabs.harvard.edu/abs/2000ApJ...539L..13G}
}

@ARTICLE{Chartas02,
   author = {{Chartas}, G. and {Brandt}, W.~N. and {Gallagher}, S.~C. and 
	{Garmire}, G.~P.},
    title = "{CHANDRA Detects Relativistic Broad Absorption Lines from APM 08279+5255}",
  journal = {\apj},
   eprint = {astro-ph/0207196},
     year = 2002,
    month = nov,
   volume = 579,
    pages = {169-175},
      doi = {10.1086/342744},
   adsurl = {http://adsabs.harvard.edu/abs/2002ApJ...579..169C}
}

@ARTICLE{Kaastra14,
   author = {{Kaastra}, J.~S. and {Kriss}, G.~A. and {Cappi}, M. and {Mehdipour}, M. and 
	{Petrucci}, P.-O. and {Steenbrugge}, K.~C. and {Arav}, N. and 
	{Behar}, E. and {Bianchi}, S. and {Boissay}, R. and {Branduardi-Raymont}, G. and 
	{Chamberlain}, C. and {Costantini}, E. and {Ely}, J.~C. and 
	{Ebrero}, J. and {Di Gesu}, L. and {Harrison}, F.~A. and {Kaspi}, S. and 
	{Malzac}, J. and {De Marco}, B. and {Matt}, G. and {Nandra}, K. and 
	{Paltani}, S. and {Person}, R. and {Peterson}, B.~M. and {Pinto}, C. and 
	{Ponti}, G. and {Nu{\~n}ez}, F.~P. and {De Rosa}, A. and {Seta}, H. and 
	{Ursini}, F. and {de Vries}, C.~P. and {Walton}, D.~J. and {Whewell}, M.
	},
    title = "{A fast and long-lived outflow from the supermassive black hole in NGC 5548}",
  journal = {Science},
archivePrefix = "arXiv",
   eprint = {1406.5007},
 primaryClass = "astro-ph.HE",
     year = 2014,
    month = jul,
   volume = 345,
    pages = {64-68},
      doi = {10.1126/science.1253787},
   adsurl = {http://adsabs.harvard.edu/abs/2014Sci...345...64K}
}

@ARTICLE{Kormendy95,
   author = {{Kormendy}, J. and {Richstone}, D.},
    title = "{Inward Bound---The Search For Supermassive Black Holes In Galactic Nuclei}",
  journal = {\araa},
     year = 1995,
   volume = 33,
    pages = {581},
      doi = {10.1146/annurev.aa.33.090195.003053},
   adsurl = {http://adsabs.harvard.edu/abs/1995ARA%26A..33..581K}
}

@ARTICLE{Magorrian98,
   author = {{Magorrian}, J. and {Tremaine}, S. and {Richstone}, D. and {Bender}, R. and 
	{Bower}, G. and {Dressler}, A. and {Faber}, S.~M. and {Gebhardt}, K. and 
	{Green}, R. and {Grillmair}, C. and {Kormendy}, J. and {Lauer}, T.
	},
    title = "{The Demography of Massive Dark Objects in Galaxy Centers}",
  journal = {\aj},
   eprint = {astro-ph/9708072},
     year = 1998,
    month = jun,
   volume = 115,
    pages = {2285-2305},
      doi = {10.1086/300353},
   adsurl = {http://adsabs.harvard.edu/abs/1998AJ....115.2285M}
}

@ARTICLE{Kormendy13,
   author = {{Kormendy}, J. and {Ho}, L.~C.},
    title = "{Coevolution (Or Not) of Supermassive Black Holes and Host Galaxies}",
  journal = {\araa},
archivePrefix = "arXiv",
   eprint = {1304.7762},
     year = 2013,
    month = aug,
   volume = 51,
    pages = {511-653},
      doi = {10.1146/annurev-astro-082708-101811},
   adsurl = {http://adsabs.harvard.edu/abs/2013ARA%26A..51..511K}
}

@ARTICLE{Kara17,
       author = {{Kara}, E. and {Garc{\'\i}a}, J.~A. and {Lohfink}, A. and {Fabian}, A.~C. and {Reynolds}, C.~S. and {Tombesi}, F. and {Wilkins}, D.~R.},
        title = "{The high-Eddington NLS1 Ark 564 has the coolest corona}",
      journal = {\mnras},
         year = 2017,
        month = jul,
       volume = {468},
       number = {3},
        pages = {3489-3498},
          doi = {10.1093/mnras/stx792},
archivePrefix = {arXiv},
       eprint = {1703.09815},
 primaryClass = {astro-ph.HE},
       adsurl = {https://ui.adsabs.harvard.edu/abs/2017MNRAS.468.3489K}
}

@ARTICLE{Laurenti21,
       author = {{Laurenti}, M. and {Luminari}, A. and {Tombesi}, F. and {Vagnetti}, F. and {Middei}, R. and {Piconcelli}, E.},
        title = "{Location and energetics of the ultra-fast outflow in PG 1448+273}",
      journal = {\aap},
         year = 2021,
        month = jan,
       volume = {645},
          eid = {A118},
        pages = {A118},
          doi = {10.1051/0004-6361/202039409},
archivePrefix = {arXiv},
       eprint = {2011.08212},
 primaryClass = {astro-ph.HE},
       adsurl = {https://ui.adsabs.harvard.edu/abs/2021A&A...645A.118L}
}

@article{Giustini11,
	Adsurl = {http://adsabs.harvard.edu/abs/2011A%26A...536A..49G},
	Archiveprefix = {arXiv},
	Author = {{Giustini}, M. and {Cappi}, M. and {Chartas}, G. and {Dadina}, M. and {Eracleous}, M. and {Ponti}, G. and {Proga}, D. and {Tombesi}, F. and {Vignali}, C. and {Palumbo}, G.~G.~C.},
	Doi = {10.1051/0004-6361/201117732},
	Eid = {A49},
	Eprint = {1109.6026},
	Journal = {\aap},
	Month = dec,
	Pages = {A49},
	Primaryclass = {astro-ph.CO},
	Title = {{Variable X-ray absorption in the mini-BAL QSO PG 1126-041}},
	Volume = 536,
	Year = 2011}

@ARTICLE{Gofford15,
   author = {{Gofford}, J. and {Reeves}, J.~N. and {McLaughlin}, D.~E. and 
	{Braito}, V. and {Turner}, T.~J. and {Tombesi}, F. and {Cappi}, M.
	},
    title = "{The Suzaku view of highly ionized outflows in AGN - II. Location, energetics and scalings with bolometric luminosity}",
  journal = {\mnras},
archivePrefix = "arXiv",
   eprint = {1506.00614},
 primaryClass = "astro-ph.HE",
     year = 2015,
    month = aug,
   volume = 451,
    pages = {4169-4182},
      doi = {10.1093/mnras/stv1207},
   adsurl = {http://adsabs.harvard.edu/abs/2015MNRAS.451.4169G}
}

@ARTICLE{Laha21,
       author = {{Laha}, Sibasish and {Reynolds}, Christopher S. and {Reeves}, James and {Kriss}, Gerard and {Guainazzi}, Matteo and {Smith}, Randall and {Veilleux}, Sylvain and {Proga}, Daniel},
        title = "{Ionized outflows from active galactic nuclei as the essential elements of feedback}",
      journal = {Nature Astronomy},
         year = 2021,
        month = jan,
       volume = {5},
        pages = {13-24},
          doi = {10.1038/s41550-020-01255-2},
archivePrefix = {arXiv},
       eprint = {2012.06945},
 primaryClass = {astro-ph.GA},
       adsurl = {https://ui.adsabs.harvard.edu/abs/2021NatAs...5...13L}
}

@ARTICLE{Zubova12,
   author = {{Zubovas}, K. and {King}, A.},
    title = "{Clearing Out a Galaxy}",
  journal = {\apjl},
archivePrefix = "arXiv",
   eprint = {1201.0866},
     year = 2012,
    month = feb,
   volume = 745,
      eid = {L34},
    pages = {L34},
      doi = {10.1088/2041-8205/745/2/L34},
   adsurl = {http://adsabs.harvard.edu/abs/2012ApJ...745L..34Z}
}

@ARTICLE{Silk98,
   author = {{Silk}, J. and {Rees}, M.~J.},
    title = "{Quasars and galaxy formation}",
  journal = {\aap},
   eprint = {astro-ph/9801013},
     year = 1998,
    month = mar,
   volume = 331,
    pages = {L1-L4},
   adsurl = {http://adsabs.harvard.edu/abs/1998A%26A...331L...1S}
}

@ARTICLE{Fukumura17,
   author = {{Fukumura}, K. and {Kazanas}, D. and {Shrader}, C. and {Behar}, E. and 
	{Tombesi}, F. and {Contopoulos}, I.},
    title = "{Magnetic origin of black hole winds across the mass scale}",
  journal = {Nature Astronomy},
archivePrefix = "arXiv",
   eprint = {1702.02197},
 primaryClass = "astro-ph.HE",
     year = 2017,
    month = mar,
   volume = 1,
      eid = {0062},
    pages = {0062},
      doi = {10.1038/s41550-017-0062},
   adsurl = {http://adsabs.harvard.edu/abs/2017NatAs...1E..62F}
}

@ARTICLE{Fiore17,
   author = {{Fiore}, F. and {Feruglio}, C. and {Shankar}, F. and {Bischetti}, M. and 
	{Bongiorno}, A. and {Brusa}, M. and {Carniani}, S. and {Cicone}, C. and 
	{Duras}, F. and {Lamastra}, A. and {Mainieri}, V. and {Marconi}, A. and 
	{Menci}, N. and {Maiolino}, R. and {Piconcelli}, E. and {Vietri}, G. and 
	{Zappacosta}, L.},
    title = "{AGN wind scaling relations and the co-evolution of black holes and galaxies}",
  journal = {\aap},
archivePrefix = "arXiv",
   eprint = {1702.04507},
     year = 2017,
    month = may,
   volume = 601,
      eid = {A143},
    pages = {A143},
      doi = {10.1051/0004-6361/201629478},
   adsurl = {http://adsabs.harvard.edu/abs/2017A%26A...601A.143F}
}

@ARTICLE{Yaqoob12,
   author = {{Yaqoob}, T.},
    title = "{The nature of the Compton-thick X-ray reprocessor in NGC 4945}",
  journal = {\mnras},
archivePrefix = "arXiv",
   eprint = {1204.4196},
 primaryClass = "astro-ph.HE",
     year = 2012,
    month = jul,
   volume = 423,
    pages = {3360-3396},
      doi = {10.1111/j.1365-2966.2012.21129.x},
   adsurl = {http://adsabs.harvard.edu/abs/2012MNRAS.423.3360Y}
}

@ARTICLE{Cash79,
   author = {{Cash}, W.},
    title = "{Parameter estimation in astronomy through application of the likelihood ratio}",
  journal = {\apj},
     year = 1979,
    month = mar,
   volume = 228,
    pages = {939-947},
      doi = {10.1086/156922},
   adsurl = {http://adsabs.harvard.edu/abs/1979ApJ...228..939C}
}

@ARTICLE{Nandra07,
   author = {{Nandra}, K. and {O'Neill}, P.~M. and {George}, I.~M. and {Reeves}, J.~N.
	},
    title = "{An XMM-Newton survey of broad iron lines in Seyfert galaxies}",
  journal = {\mnras},
archivePrefix = "arXiv",
   eprint = {0708.1305},
     year = 2007,
    month = nov,
   volume = 382,
    pages = {194-228},
      doi = {10.1111/j.1365-2966.2007.12331.x},
   adsurl = {http://adsabs.harvard.edu/abs/2007MNRAS.382..194N}
}

@ARTICLE{Balokovic18,
   author = {{Balokovi{\'c}}, M. and {Brightman}, M. and {Harrison}, F.~A. and 
	{Comastri}, A. and {Ricci}, C. and {Buchner}, J. and {Gandhi}, P. and 
	{Farrah}, D. and {Stern}, D.},
    title = "{New Spectral Model for Constraining Torus Covering Factors from Broadband X-Ray Spectra of Active Galactic Nuclei}",
  journal = {\apj},
archivePrefix = "arXiv",
   eprint = {1801.04938},
 primaryClass = "astro-ph.HE",
     year = 2018,
    month = feb,
   volume = 854,
      eid = {42},
    pages = {42},
      doi = {10.3847/1538-4357/aaa7eb},
   adsurl = {https://ui.adsabs.harvard.edu/abs/2018ApJ...854...42B}
}

@ARTICLE{Ponti12,
   author = {{Ponti}, G. and {Papadakis}, I. and {Bianchi}, S. and {Guainazzi}, M. and 
	{Matt}, G. and {Uttley}, P. and {Bonilla}, N.~F.},
    title = "{CAIXA: a catalogue of AGN in the XMM-Newton archive. III. Excess variance analysis}",
  journal = {\aap},
archivePrefix = "arXiv",
   eprint = {1112.2744},
 primaryClass = "astro-ph.HE",
     year = 2012,
    month = jun,
   volume = 542,
      eid = {A83},
    pages = {A83},
      doi = {10.1051/0004-6361/201118326},
   adsurl = {https://ui.adsabs.harvard.edu/abs/2012A%26A...542A..83P}
}

@ARTICLE{Igo20,
       author = {{Igo}, Z. and {Parker}, M.~L. and {Matzeu}, G.~A. and {Alston}, W. and
         {Alvarez Crespo}, N. and {F{\"u}rst}, F. and {Buisson}, D.~J.~K. and
         {Lobban}, A. and {Joyce}, A.~M. and {Mallick}, L. and {Schartel}, N. and
         {Santos-Lle{\'o}}, M.},
        title = "{Searching for ultra-fast outflows in AGN using variability spectra}",
      journal = {\mnras},
         year = 2020,
        month = mar,
       volume = {493},
       number = {1},
        pages = {1088-1108},
          doi = {10.1093/mnras/staa265},
archivePrefix = {arXiv},
       eprint = {2001.08208},
 primaryClass = {astro-ph.HE},
       adsurl = {https://ui.adsabs.harvard.edu/abs/2020MNRAS.493.1088I}
}

@ARTICLE{KaastraBleeker16,
       author = {{Kaastra}, J.~S. and {Bleeker}, J.~A.~M.},
        title = "{Optimal binning of X-ray spectra and response matrix design}",
      journal = {\aap},
         year = 2016,
        month = mar,
       volume = {587},
          eid = {A151},
        pages = {A151},
          doi = {10.1051/0004-6361/201527395},
archivePrefix = {arXiv},
       eprint = {1601.05309},
 primaryClass = {astro-ph.IM},
       adsurl = {https://ui.adsabs.harvard.edu/abs/2016A&A...587A.151K}
}

@ARTICLE{Luminari21,
       author = {{Luminari}, A. and {Nicastro}, F. and {Elvis}, M. and {Piconcelli}, E. and {Tombesi}, F. and {Zappacosta}, L. and {Fiore}, F.},
        title = "{Speed limits for radiation-driven SMBH winds}",
      journal = {\aap},
         year = 2021,
        month = feb,
       volume = {646},
          eid = {A111},
        pages = {A111},
          doi = {10.1051/0004-6361/202039396},
archivePrefix = {arXiv},
       eprint = {2012.07877},
 primaryClass = {astro-ph.HE},
       adsurl = {https://ui.adsabs.harvard.edu/abs/2021A&A...646A.111L}
}

@ARTICLE{Luminari20,
       author = {{Luminari}, A. and {Tombesi}, F. and {Piconcelli}, E. and {Nicastro}, F. and {Fukumura}, K. and {Kazanas}, D. and {Fiore}, F. and {Zappacosta}, L.},
        title = "{The importance of special relativistic effects in modelling ultra-fast outflows}",
      journal = {\aap},
         year = 2020,
        month = jan,
       volume = {633},
          eid = {A55},
        pages = {A55},
          doi = {10.1051/0004-6361/201936797},
archivePrefix = {arXiv},
       eprint = {1912.00494},
 primaryClass = {astro-ph.HE},
       adsurl = {https://ui.adsabs.harvard.edu/abs/2020A&A...633A..55L}
}

@ARTICLE{ReevesBraito19,
       author = {{Reeves}, J.~N. and {Braito}, V.},
        title = "{A Momentum-conserving Accretion Disk Wind in the Narrow-line Seyfert 1 I Zwicky 1}",
      journal = {\apj},
         year = 2019,
        month = oct,
       volume = {884},
       number = {1},
          eid = {80},
        pages = {80},
          doi = {10.3847/1538-4357/ab41f9},
archivePrefix = {arXiv},
       eprint = {1909.05039},
 primaryClass = {astro-ph.GA},
       adsurl = {https://ui.adsabs.harvard.edu/abs/2019ApJ...884...80R}
}

@ARTICLE{Miniutti06,
       author = {{Miniutti}, G. and {Fabian}, A.~C.},
        title = "{Discovery of a relativistic Fe line in PG 1425+267 with XMM-Newton and study of its short time-scale variability}",
      journal = {\mnras},
         year = 2006,
        month = feb,
       volume = {366},
       number = {1},
        pages = {115-124},
          doi = {10.1111/j.1365-2966.2005.09805.x},
archivePrefix = {arXiv},
       eprint = {astro-ph/0510858},
 primaryClass = {astro-ph},
       adsurl = {https://ui.adsabs.harvard.edu/abs/2006MNRAS.366..115M}
}

@ARTICLE{KingPounds15,
       author = {{King}, Andrew and {Pounds}, Ken},
        title = "{Powerful Outflows and Feedback from Active Galactic Nuclei}",
      journal = {\araa},
         year = 2015,
        month = aug,
       volume = {53},
        pages = {115-154},
          doi = {10.1146/annurev-astro-082214-122316},
archivePrefix = {arXiv},
       eprint = {1503.05206},
 primaryClass = {astro-ph.GA},
       adsurl = {https://ui.adsabs.harvard.edu/abs/2015ARA&A..53..115K}
}

@ARTICLE{Chartas21,
       author = {{Chartas}, G. and {Cappi}, M. and {Vignali}, C. and {Dadina}, M. and {James}, V. and {Lanzuisi}, G. and {Giustini}, M. and {Gaspari}, M. and {Strickland}, S. and {Bertola}, E.},
        title = "{Multiphase Powerful Outflows Detected in High-z Quasars}",
      journal = {\apj},
         year = 2021,
        month = oct,
       volume = {920},
       number = {1},
          eid = {24},
        pages = {24},
          doi = {10.3847/1538-4357/ac0ef2},
archivePrefix = {arXiv},
       eprint = {2106.14907},
 primaryClass = {astro-ph.GA},
       adsurl = {https://ui.adsabs.harvard.edu/abs/2021ApJ...920...24C}
}

@ARTICLE{Chartas18IRAS13224,
       author = {{Chartas}, George and {Canas}, Manuel H.},
        title = "{The Variable Relativistic Outflow of IRAS 13224-3809}",
      journal = {\apj},
         year = 2018,
        month = nov,
       volume = {867},
       number = {2},
          eid = {103},
        pages = {103},
          doi = {10.3847/1538-4357/aae438},
archivePrefix = {arXiv},
       eprint = {1809.09138},
 primaryClass = {astro-ph.HE},
       adsurl = {https://ui.adsabs.harvard.edu/abs/2018ApJ...867..103C}
}

@ARTICLE{Kraemer18,
       author = {{Kraemer}, S.~B. and {Tombesi}, F. and {Bottorff}, M.~C.},
        title = "{Physical Conditions in Ultra-fast Outflows in AGN}",
      journal = {\apj},
         year = 2018,
        month = jan,
       volume = {852},
       number = {1},
          eid = {35},
        pages = {35},
          doi = {10.3847/1538-4357/aa9ce0},
archivePrefix = {arXiv},
       eprint = {1711.07965},
 primaryClass = {astro-ph.GA},
       adsurl = {https://ui.adsabs.harvard.edu/abs/2018ApJ...852...35K}
}

@ARTICLE{Matzeu23,
       author = {{Matzeu}, G.~A. and {Brusa}, M. and {Lanzuisi}, G. and {Dadina}, M. and {Bianchi}, S. and {Kriss}, G. and {Mehdipour}, M. and {Nardini}, E. and {Chartas}, G. and {Middei}, R. and {Piconcelli}, E. and {Gianolli}, V. and {Comastri}, A. and {Longinotti}, A.~L. and {Krongold}, Y. and {Ricci}, F. and {Petrucci}, P.~O. and {Tombesi}, F. and {Luminari}, A. and {Zappacosta}, L. and {Miniutti}, G. and {Gaspari}, M. and {Behar}, E. and {Bischetti}, M. and {Mathur}, S. and {Perna}, M. and {Giustini}, M. and {Grandi}, P. and {Torresi}, E. and {Vignali}, C. and {Bruni}, G. and {Cappi}, M. and {Costantini}, E. and {Cresci}, G. and {De Marco}, B. and {De Rosa}, A. and {Gilli}, R. and {Guainazzi}, M. and {Kaastra}, J. and {Kraemer}, S. and {La Franca}, F. and {Marconi}, A. and {Panessa}, F. and {Ponti}, G. and {Proga}, D. and {Ursini}, F. and {Baldini}, P. and {Fiore}, F. and {King}, A.~R. and {Maiolino}, R. and {Matt}, G. and {Merloni}, A.},
        title = "{Supermassive Black Hole Winds in X-rays: SUBWAYS. I. Ultra-fast outflows in quasars beyond the local Universe}",
      journal = {\aap},
         year = 2023,
        month = feb,
       volume = {670},
          eid = {A182},
        pages = {A182},
          doi = {10.1051/0004-6361/202245036},
archivePrefix = {arXiv},
       eprint = {2212.02960},
 primaryClass = {astro-ph.HE},
       adsurl = {https://ui.adsabs.harvard.edu/abs/2023A&A...670A.182M}
}

@ARTICLE{Fukumura22,
       author = {{Fukumura}, Keigo and {Dadina}, Mauro and {Matzeu}, Gabriele and {Tombesi}, Francesco and {Shrader}, Chris and {Kazanas}, Demosthenes},
        title = "{Tell-tale Spectral Signatures of MHD-driven Ultrafast Outflows in AGNs}",
      journal = {\apj},
         year = 2022,
        month = nov,
       volume = {940},
       number = {1},
          eid = {6},
        pages = {6},
          doi = {10.3847/1538-4357/ac9388},
archivePrefix = {arXiv},
       eprint = {2205.08894},
 primaryClass = {astro-ph.HE},
       adsurl = {https://ui.adsabs.harvard.edu/abs/2022ApJ...940....6F}
}

@article{Lanzuisi26sub,
  title={},
  author={Lanzuisi, G. and Borrelli, L. and Piconcelli, E. and Brusa, M. and Comastri A. and},
  journal={Submitted to A\&A},
  volume={},
  number={},
  pages={},
  year={2026},
  publisher={Oxford University Press}
}

@ARTICLE{Svoboda17,
       author = {{Svoboda}, J. and {Guainazzi}, M. and {Merloni}, A.},
        title = "{AGN spectral states from simultaneous UV and X-ray observations by XMM-Newton}",
      journal = {\aap},
         year = 2017,
        month = jul,
       volume = {603},
          eid = {A127},
        pages = {A127},
          doi = {10.1051/0004-6361/201630181},
archivePrefix = {arXiv},
       eprint = {1704.07268},
 primaryClass = {astro-ph.GA},
       adsurl = {https://ui.adsabs.harvard.edu/abs/2017A&A...603A.127S}
}

@ARTICLE{Magdziarz95,
       author = {{Magdziarz}, Pawel and {Zdziarski}, Andrzej A.},
        title = "{Angle-dependent Compton reflection of X-rays and gamma-rays}",
      journal = {\mnras},
         year = 1995,
        month = apr,
       volume = {273},
       number = {3},
        pages = {837-848},
          doi = {10.1093/mnras/273.3.837},
       adsurl = {https://ui.adsabs.harvard.edu/abs/1995MNRAS.273..837M}
}

@ARTICLE{Reeves18A,
       author = {{Reeves}, J.~N. and {Braito}, V. and {Nardini}, E. and {Lobban}, A.~P. and {Matzeu}, G.~A. and {Costa}, M.~T.},
        title = "{A New Relativistic Component of the Accretion Disk Wind in PDS 456}",
      journal = {\apjl},
         year = 2018,
        month = feb,
       volume = {854},
       number = {1},
          eid = {L8},
        pages = {L8},
          doi = {10.3847/2041-8213/aaaae1},
archivePrefix = {arXiv},
       eprint = {1801.08899},
 primaryClass = {astro-ph.HE},
       adsurl = {https://ui.adsabs.harvard.edu/abs/2018ApJ...854L...8R}
}

@ARTICLE{Reeves18B,
       author = {{Reeves}, J.~N. and {Braito}, V. and {Nardini}, E. and {Hamann}, F. and {Chartas}, G. and {Lobban}, A.~P. and {O'Brien}, P.~T. and {Turner}, T.~J.},
        title = "{Resolving the X-Ray Obscuration in a Low-flux Observation of the Quasar PDS 456}",
      journal = {\apj},
         year = 2018,
        month = nov,
       volume = {867},
       number = {1},
          eid = {38},
        pages = {38},
          doi = {10.3847/1538-4357/aae30c},
archivePrefix = {arXiv},
       eprint = {1809.07164},
 primaryClass = {astro-ph.HE},
       adsurl = {https://ui.adsabs.harvard.edu/abs/2018ApJ...867...38R}
}

@ARTICLE{Luminari23,
       author = {{Luminari}, Alfredo and {Marinucci}, Andrea and {Bianchi}, Stefano and {de Marco}, Barbara and {Feruglio}, Chiara and {Matt}, Giorgio and {Middei}, Riccardo and {Nardini}, Emanuele and {Piconcelli}, Enrico and {Puccetti}, Simonetta and {Tombesi}, Francesco},
        title = "{The Lively Accretion Disk in NGC 2992. III. Tentative Evidence of Rapid Ultrafast Outflow Variability}",
      journal = {\apj},
         year = 2023,
        month = jun,
       volume = {950},
       number = {2},
          eid = {160},
        pages = {160},
          doi = {10.3847/1538-4357/acd2d8},
archivePrefix = {arXiv},
       eprint = {2305.03754},
 primaryClass = {astro-ph.GA},
       adsurl = {https://ui.adsabs.harvard.edu/abs/2023ApJ...950..160L}
}

@ARTICLE{Bentz18,
       author = {{Bentz}, Misty C. and {Manne-Nicholas}, Emily},
        title = "{Black Hole-Galaxy Scaling Relationships for Active Galactic Nuclei with Reverberation Masses}",
      journal = {\apj},
         year = 2018,
        month = sep,
       volume = {864},
       number = {2},
          eid = {146},
        pages = {146},
          doi = {10.3847/1538-4357/aad808},
archivePrefix = {arXiv},
       eprint = {1808.01329},
 primaryClass = {astro-ph.GA},
       adsurl = {https://ui.adsabs.harvard.edu/abs/2018ApJ...864..146B}
}

@ARTICLE{Fausnaugh17,
       author = {{Fausnaugh}, Michael M. and {Peterson}, Bradley M. and {Starkey}, David A. and {Horne}, Keith and {AGN Storm Collaboration}},
        title = "{Continuum Reverberation Mapping of AGN Accretion Disks}",
      journal = {Frontiers in Astronomy and Space Sciences},
         year = 2017,
        month = dec,
       volume = {4},
          eid = {55},
        pages = {55},
          doi = {10.3389/fspas.2017.00055},
       adsurl = {https://ui.adsabs.harvard.edu/abs/2017FrASS...4...55F}
}

@ARTICLE{Bentz15,
       author = {{Bentz}, Misty C. and {Katz}, Sarah},
        title = "{The AGN Black Hole Mass Database}",
      journal = {\pasp},
         year = 2015,
        month = jan,
       volume = {127},
       number = {947},
        pages = {67},
          doi = {10.1086/679601},
archivePrefix = {arXiv},
       eprint = {1411.2596},
 primaryClass = {astro-ph.GA},
       adsurl = {https://ui.adsabs.harvard.edu/abs/2015PASP..127...67B}
}

@ARTICLE{Vestergaard06,
       author = {{Vestergaard}, Marianne and {Peterson}, Bradley M.},
        title = "{Determining Central Black Hole Masses in Distant Active Galaxies and Quasars. II. Improved Optical and UV Scaling Relationships}",
      journal = {\apj},
         year = 2006,
        month = apr,
       volume = {641},
       number = {2},
        pages = {689-709},
          doi = {10.1086/500572},
archivePrefix = {arXiv},
       eprint = {astro-ph/0601303},
 primaryClass = {astro-ph},
       adsurl = {https://ui.adsabs.harvard.edu/abs/2006ApJ...641..689V}
}

@ARTICLE{Mej22,
       author = {{Mej{\'\i}a-Restrepo}, Julian E. and {Trakhtenbrot}, Benny and {Koss}, Michael J. and {Oh}, Kyuseok and {den Brok}, Jakob and {Stern}, Daniel and {Powell}, Meredith C. and {Ricci}, Federica and {Caglar}, Turgay and {Ricci}, Claudio and {Bauer}, Franz E. and {Treister}, Ezequiel and {Harrison}, Fiona A. and {Urry}, C.~M. and {Ananna}, Tonima Tasnim and {Asmus}, Daniel and {Assef}, Roberto J. and {B{\"a}r}, Rudolf E. and {Bessiere}, Patricia S. and {Burtscher}, Leonard and {Ichikawa}, Kohei and {Kakkad}, Darshan and {Kamraj}, Nikita and {Mushotzky}, Richard and {Privon}, George C. and {Rojas}, Alejandra F. and {Sani}, Eleonora and {Schawinski}, Kevin and {Veilleux}, Sylvain},
        title = "{BASS. XXV. DR2 Broad-line-based Black Hole Mass Estimates and Biases from Obscuration}",
      journal = {\apjs},
         year = 2022,
        month = jul,
       volume = {261},
       number = {1},
          eid = {5},
        pages = {5},
          doi = {10.3847/1538-4365/ac6602},
archivePrefix = {arXiv},
       eprint = {2204.05321},
 primaryClass = {astro-ph.GA},
       adsurl = {https://ui.adsabs.harvard.edu/abs/2022ApJS..261....5M}
}

@ARTICLE{Robinson21,
       author = {{Robinson}, Justin H. and {Bentz}, Misty C. and {Courtois}, H{\'e}l{\`e}ne M. and {Johnson}, Megan C. and {Crenshaw}, D.~M. and {Meena}, Beena and {Polack}, Garrett E. and {Silverstein}, Michele L. and {Chen}, Dading},
        title = "{Tully-Fisher Distances and Dynamical Mass Constraints for 24 Host Galaxies of Reverberation-mapped AGNs}",
      journal = {\apj},
         year = 2021,
        month = may,
       volume = {912},
       number = {2},
          eid = {160},
        pages = {160},
          doi = {10.3847/1538-4357/abedaa},
archivePrefix = {arXiv},
       eprint = {2103.07000},
 primaryClass = {astro-ph.GA},
       adsurl = {https://ui.adsabs.harvard.edu/abs/2021ApJ...912..160R}
}

@ARTICLE{Humphrey09,
       author = {{Humphrey}, Philip J. and {Liu}, Wenhao and {Buote}, David A.},
        title = "{{\ensuremath{\chi}}$^{2}$ and Poissonian Data: Biases Even in the High-Count Regime and How to Avoid Them}",
      journal = {\apj},
         year = 2009,
        month = mar,
       volume = {693},
       number = {1},
        pages = {822-829},
          doi = {10.1088/0004-637X/693/1/822},
archivePrefix = {arXiv},
       eprint = {0811.2796},
 primaryClass = {astro-ph},
       adsurl = {https://ui.adsabs.harvard.edu/abs/2009ApJ...693..822H}
}

@ARTICLE{Kelly07,
       author = {{Kelly}, Brandon C.},
        title = "{Some Aspects of Measurement Error in Linear Regression of Astronomical Data}",
      journal = {\apj},
         year = 2007,
        month = aug,
       volume = {665},
       number = {2},
        pages = {1489-1506},
          doi = {10.1086/519947},
archivePrefix = {arXiv},
       eprint = {0705.2774},
 primaryClass = {astro-ph},
       adsurl = {https://ui.adsabs.harvard.edu/abs/2007ApJ...665.1489K}
}

@article{Kaastra_2017,
   title={On the use of C-stat in testing models for X-ray spectra},
   volume={605},
   ISSN={1432-0746},
   url={http://dx.doi.org/10.1051/0004-6361/201629319},
   DOI={10.1051/0004-6361/201629319},
   journal={\aap},
   publisher={EDP Sciences},
   author={Kaastra, J. S.},
   year={2017},
   month=sep, pages={A51} }

@ARTICLE{Lampton76,
       author = {{Lampton}, M. and {Margon}, B. and {Bowyer}, S.},
        title = "{Parameter estimation in X-ray astronomy.}",
      journal = {\apj},
         year = 1976,
        month = aug,
       volume = {208},
        pages = {177-190},
          doi = {10.1086/154592},
       adsurl = {https://ui.adsabs.harvard.edu/abs/1976ApJ...208..177L}
}

@INCOLLECTION{Schartel_2022,
       author = {{Schartel}, Norbert and {Gonz{\'a}lez-Riestra}, Rosario and {Kretschmar}, Peter and {Kirsch}, Marcus and {Rodr{\'\i}guez-Pascual}, Pedro and {Rosen}, Simon and {Santos-Lle{\'o}}, Maria and {Smith}, Michael and {Stuhlinger}, Martin and {Verdugo-Rodrigo}, Eva},
        title = "{XMM-Newton}",
    booktitle = {Handbook of X-ray and Gamma-ray Astrophysics},
         year = 2022,
       editor = {{Bambi}, Cosimo and {Sangangelo}, Andrea},
          eid = {114},
        pages = {114},
          doi = {10.1007/978-981-16-4544-0_41-1},
       adsurl = {https://ui.adsabs.harvard.edu/abs/2022hxga.book..114S}
}

@ARTICLE{Murphy09,
       author = {{Murphy}, Kendrah D. and {Yaqoob}, Tahir},
        title = "{An X-ray spectral model for Compton-thick toroidal reprocessors}",
      journal = {\mnras},
         year = 2009,
        month = aug,
       volume = {397},
       number = {3},
        pages = {1549-1562},
          doi = {10.1111/j.1365-2966.2009.15025.x},
archivePrefix = {arXiv},
       eprint = {0905.3188},
 primaryClass = {astro-ph.HE},
       adsurl = {https://ui.adsabs.harvard.edu/abs/2009MNRAS.397.1549M}
}

@ARTICLE{Jana21,
       author = {{Jana}, Arghajit and {Kumari}, Neeraj and {Nandi}, Prantik and {Naik}, Sachindra and {Chatterjee}, Arka and {Jaisawal}, Gaurava K. and {Hayasaki}, Kimitake and {Ricci}, Claudio},
        title = "{Broad-band X-ray observations of the 2018 outburst of the changing-look active galactic nucleus NGC 1566}",
      journal = {\mnras},
         year = 2021,
        month = oct,
       volume = {507},
       number = {1},
        pages = {687-703},
          doi = {10.1093/mnras/stab2155},
archivePrefix = {arXiv},
       eprint = {2107.11127},
 primaryClass = {astro-ph.HE},
       adsurl = {https://ui.adsabs.harvard.edu/abs/2021MNRAS.507..687J}
}

@ARTICLE{Giustini17,
       author = {{Giustini}, M. and {Costantini}, E. and {De Marco}, B. and {Svoboda}, J. and {Motta}, S.~E. and {Proga}, D. and {Saxton}, R. and {Ferrigno}, C. and {Longinotti}, A.~L. and {Miniutti}, G. and {Grupe}, D. and {Mathur}, S. and {Shappee}, B.~J. and {Prieto}, J.~L. and {Stanek}, K.},
        title = "{Direct probe of the inner accretion flow around the supermassive black hole in NGC 2617}",
      journal = {\aap},
         year = 2017,
        month = jan,
       volume = {597},
          eid = {A66},
        pages = {A66},
          doi = {10.1051/0004-6361/201628686},
archivePrefix = {arXiv},
       eprint = {1608.00233},
 primaryClass = {astro-ph.GA},
       adsurl = {https://ui.adsabs.harvard.edu/abs/2017A&A...597A..66G}
}

@ARTICLE{Tombesi14,
       author = {{Tombesi}, F. and {Tazaki}, F. and {Mushotzky}, R.~F. and {Ueda}, Y. and {Cappi}, M. and {Gofford}, J. and {Reeves}, J.~N. and {Guainazzi}, M.},
        title = "{Ultrafast outflows in radio-loud active galactic nuclei}",
      journal = {\mnras},
         year = 2014,
        month = sep,
       volume = {443},
       number = {3},
        pages = {2154-2182},
          doi = {10.1093/mnras/stu1297},
archivePrefix = {arXiv},
       eprint = {1406.7252},
 primaryClass = {astro-ph.HE},
       adsurl = {https://ui.adsabs.harvard.edu/abs/2014MNRAS.443.2154T}
}

@ARTICLE{Madsen15,
       author = {{Madsen}, Kristin K. and {F{\"u}rst}, Felix and {Walton}, Dominic J. and {Harrison}, Fiona A. and {Nalewajko}, Krzysztof and {Ballantyne}, David R. and {Boggs}, Steve E. and {Brenneman}, Laura W. and {Christensen}, Finn E. and {Craig}, William W. and {Fabian}, Andrew C. and {Forster}, Karl and {Grefenstette}, Brian W. and {Guainazzi}, Matteo and {Hailey}, Charles J. and {Madejski}, Greg M. and {Matt}, Giorgio and {Stern}, Daniel and {Walter}, Roland and {Zhang}, William W.},
        title = "{3C 273 with NuSTAR: Unveiling the Active Galactic Nucleus}",
      journal = {\apj},
         year = 2015,
        month = oct,
       volume = {812},
       number = {1},
          eid = {14},
        pages = {14},
          doi = {10.1088/0004-637X/812/1/14},
archivePrefix = {arXiv},
       eprint = {1506.06182},
 primaryClass = {astro-ph.GA},
       adsurl = {https://ui.adsabs.harvard.edu/abs/2015ApJ...812...14M}
}

@ARTICLE{carter07,
       author = {{Carter}, J.~A. and {Read}, A.~M.},
        title = "{The XMM-Newton EPIC background and the production of background blank sky event files}",
      journal = {\aap},
         year = 2007,
        month = mar,
       volume = {464},
       number = {3},
        pages = {1155-1166},
          doi = {10.1051/0004-6361:20065882},
archivePrefix = {arXiv},
       eprint = {astro-ph/0701209},
 primaryClass = {astro-ph},
       adsurl = {https://ui.adsabs.harvard.edu/abs/2007A&A...464.1155C}
}

@ARTICLE{Zastrocky2024,
       author = {{Zastrocky}, T.~E. and {Brotherton}, Michael S. and {Du}, Pu and {McLane}, Jacob N. and {Olson}, Kianna A. and {Dale}, D.~A. and {Kobulnicky}, H.~A. and {Maithil}, Jaya and {Nguyen}, My L. and {Chick}, William T. and {Kasper}, David H. and {Hand}, Derek and {Adelman}, C. and {Carter}, Z. and {Murphree}, G. and {Oeur}, M. and {Roth}, T. and {Schonsberg}, S. and {Caradonna}, M.~J. and {Favro}, J. and {Ferguson}, A.~J. and {Gonzalez}, I.~M. and {Hadding}, L.~M. and {Hagler}, H.~D. and {Rogers}, C.~J. and {Stack}, T.~R. and {Chapman}, Franklin and {Bao}, Dong-Wei and {Fang}, Feng-Na and {Zhai}, Shuo and {Yang}, Sen and {Chen}, Yong-Jie and {Bai}, Hua-Rui and {Fu}, Yi-Xin and {Liu}, Jun-Rong and {Yao}, Zhu-Heng and {Peng}, Yue-Chang and {Songsheng}, Yu-Yang and {Li}, Yan-Rong and {Bai}, Jin-Ming and {Hu}, Chen and {Xiao}, Ming and {Ho}, Luis C. and {Wang}, Jian-Min},
        title = "{Monitoring AGNs with H{\ensuremath{\beta}} Asymmetry. IV. First Reverberation Mapping Results of 14 Active Galactic Nuclei}",
      journal = {\apjs},
         year = 2024,
        month = jun,
       volume = {272},
       number = {2},
          eid = {29},
        pages = {29},
          doi = {10.3847/1538-4365/ad3bad},
archivePrefix = {arXiv},
       eprint = {2404.07343},
 primaryClass = {astro-ph.GA},
       adsurl = {https://ui.adsabs.harvard.edu/abs/2024ApJS..272...29Z}
}

@ARTICLE{Du15,
       author = {{Du}, Pu and {Hu}, Chen and {Lu}, Kai-Xing and {Huang}, Ying-Ke and {Cheng}, Cheng and {Qiu}, Jie and {Li}, Yan-Rong and {Zhang}, Yang-Wei and {Fan}, Xu-Liang and {Bai}, Jin-Ming and {Bian}, Wei-Hao and {Yuan}, Ye-Fei and {Kaspi}, Shai and {Ho}, Luis C. and {Netzer}, Hagai and {Wang}, Jian-Min and {SEAMBH Collaboration}},
        title = "{Supermassive Black Holes with High Accretion Rates in Active Galactic Nuclei. IV. H{\ensuremath{\beta}} Time Lags and Implications for Super-Eddington Accretion}",
      journal = {\apj},
         year = 2015,
        month = jun,
       volume = {806},
       number = {1},
          eid = {22},
        pages = {22},
          doi = {10.1088/0004-637X/806/1/22},
archivePrefix = {arXiv},
       eprint = {1504.01844},
 primaryClass = {astro-ph.GA},
       adsurl = {https://ui.adsabs.harvard.edu/abs/2015ApJ...806...22D}
}

@ARTICLE{Xu2023,
       author = {{Xu}, Yerong and {Pinto}, Ciro and {Rogantini}, Daniele and {Bianchi}, Stefano and {Guainazzi}, Matteo and {Kara}, Erin and {Jin}, Chichuan and {Cusumano}, Giancarlo},
        title = "{Constraints on the ultrafast outflows in the narrow-line Seyfert 1 galaxy Mrk 1044 from high-resolution time- and flux-resolved spectroscopy}",
      journal = {\mnras},
         year = 2023,
        month = aug,
       volume = {523},
       number = {2},
        pages = {2158-2171},
          doi = {10.1093/mnras/stad1565},
archivePrefix = {arXiv},
       eprint = {2305.11966},
 primaryClass = {astro-ph.HE},
       adsurl = {https://ui.adsabs.harvard.edu/abs/2023MNRAS.523.2158X}
}

@ARTICLE{Botte2004,
       author = {{Botte}, V. and {Ciroi}, S. and {Rafanelli}, P. and {Di Mille}, F.},
        title = "{Exploring Narrow-Line Seyfert 1 Galaxies through the Physical Properties of Their Hosts}",
      journal = {\aj},
         year = 2004,
        month = jun,
       volume = {127},
       number = {6},
        pages = {3168-3179},
          doi = {10.1086/420803},
archivePrefix = {arXiv},
       eprint = {astro-ph/0402627},
 primaryClass = {astro-ph},
       adsurl = {https://ui.adsabs.harvard.edu/abs/2004AJ....127.3168B}
}

@ARTICLE{Yamada2024,
       author = {{Yamada}, Satoshi and {Kawamuro}, Taiki and {Mizumoto}, Misaki and {Ricci}, Claudio and {Ogawa}, Shoji and {Noda}, Hirofumi and {Ueda}, Yoshihiro and {Enoto}, Teruaki and {Kokubo}, Mitsuru and {Minezaki}, Takeo and {Sameshima}, Hiroaki and {Horiuchi}, Takashi and {Mizukoshi}, Shoichiro},
        title = "{X-Ray Winds in Nearby-to-distant Galaxies (X-WING). I. Legacy Surveys of Galaxies with Ultrafast Outflows and Warm Absorbers in z {\ensuremath{\sim}} 0{\textendash}4}",
      journal = {\apjs},
         year = 2024,
        month = sep,
       volume = {274},
       number = {1},
          eid = {8},
        pages = {8},
          doi = {10.3847/1538-4365/ad5961},
archivePrefix = {arXiv},
       eprint = {2405.02391},
 primaryClass = {astro-ph.HE},
       adsurl = {https://ui.adsabs.harvard.edu/abs/2024ApJS..274....8Y}
}

@ARTICLE{Behar2017,
       author = {{Behar}, Ehud and {Peretz}, Uria and {Kriss}, Gerard A. and {Kaastra}, Jelle and {Arav}, Nahum and {Bianchi}, Stefano and {Branduardi-Raymont}, Graziella and {Cappi}, Massimo and {Costantini}, Elisa and {De Marco}, Barbara and {Di Gesu}, Laura and {Ebrero}, Jacobo and {Kaspi}, Shai and {Mehdipour}, Missagh and {Paltani}, St{\'e}phane and {Petrucci}, Pierre-Olivier and {Ponti}, Gabriele and {Ursini}, Francesco},
        title = "{Multi-wavelength campaign on NGC 7469. I. The rich 640 ks RGS spectrum}",
      journal = {\aap},
         year = 2017,
        month = may,
       volume = {601},
          eid = {A17},
        pages = {A17},
          doi = {10.1051/0004-6361/201629943},
archivePrefix = {arXiv},
       eprint = {1612.07508},
 primaryClass = {astro-ph.HE},
       adsurl = {https://ui.adsabs.harvard.edu/abs/2017A&A...601A..17B}
}

@ARTICLE{Buchner16,
       author = {{Buchner}, J. and {Georgakakis}, A. and {Nandra}, K. and {Hsu}, L. and {Rangel}, C. and {Brightman}, M. and {Merloni}, A. and {Salvato}, M. and {Donley}, J. and {Kocevski}, D.},
        title = "{X-ray spectral modelling of the AGN obscuring region in the CDFS: Bayesian model selection and catalogue}",
      journal = {\aap},
         year = 2014,
        month = apr,
       volume = {564},
          eid = {A125},
        pages = {A125},
          doi = {10.1051/0004-6361/201322971},
archivePrefix = {arXiv},
       eprint = {1402.0004},
 primaryClass = {astro-ph.HE},
       adsurl = {https://ui.adsabs.harvard.edu/abs/2014A&A...564A.125B}
}

@ARTICLE{Feng2021,
       author = {{Feng}, Hai-Cheng and {Liu}, H.~T. and {Bai}, J.~M. and {Yang}, Zi-Xu and {Hu}, Chen and {Li}, Sha-Sha and {Yang}, Sen and {Lu}, Kai-Xing and {Xiao}, Ming},
        title = "{Velocity-resolved Reverberation Mapping of Changing-look AGN NGC 2617}",
      journal = {\apj},
         year = 2021,
        month = may,
       volume = {912},
       number = {2},
          eid = {92},
        pages = {92},
          doi = {10.3847/1538-4357/abefe0},
archivePrefix = {arXiv},
       eprint = {2103.03508},
 primaryClass = {astro-ph.GA},
       adsurl = {https://ui.adsabs.harvard.edu/abs/2021ApJ...912...92F}
}

@ARTICLE{Gianolli2024,
       author = {{Gianolli}, V.~E. and {Bianchi}, S. and {Petrucci}, P. -O. and {Brusa}, M. and {Chartas}, G. and {Lanzuisi}, G. and {Matzeu}, G.~A. and {Parra}, M. and {Ursini}, F. and {Behar}, E. and {Bischetti}, M. and {Comastri}, A. and {Costantini}, E. and {Cresci}, G. and {Dadina}, M. and {De Marco}, B. and {De Rosa}, A. and {Fiore}, F. and {Gaspari}, M. and {Gilli}, R. and {Giustini}, M. and {Guainazzi}, M. and {King}, A.~R. and {Kraemer}, S. and {Kriss}, G. and {Krongold}, Y. and {La Franca}, F. and {Longinotti}, A.~L. and {Luminari}, A. and {Maiolino}, R. and {Marconi}, A. and {Mathur}, S. and {Matt}, G. and {Mehdipour}, M. and {Merloni}, A. and {Middei}, R. and {Miniutti}, G. and {Nardini}, E. and {Panessa}, F. and {Perna}, M. and {Piconcelli}, E. and {Ponti}, G. and {Ricci}, F. and {Serafinelli}, R. and {Tombesi}, F. and {Vignali}, C. and {Zappacosta}, L.},
        title = "{Supermassive Black Hole Winds in X-rays: SUBWAYS. III. A population study on ultra-fast outflows}",
      journal = {\aap},
         year = 2024,
        month = jul,
       volume = {687},
          eid = {A235},
        pages = {A235},
          doi = {10.1051/0004-6361/202348908},
archivePrefix = {arXiv},
       eprint = {2403.09538},
 primaryClass = {astro-ph.GA},
       adsurl = {https://ui.adsabs.harvard.edu/abs/2024A&A...687A.235G}
}

@ARTICLE{Xu2025_PDS,
       author = {{Xu}, Yerong and {Gallo}, Luigi C and {Hagino}, Kouichi and {Reeves}, James N and {Tombesi}, Francesco and {Mizumoto}, Misaki and {Luminari}, Alfredo and {Gonzalez}, Adam G and {Behar}, Ehud and {Boissay-Malaquin}, Rozenn and {Braito}, Valentina and {Cond{\'o}}, Pierpaolo and {Done}, Chris and {Miyamoto}, Aiko and {Mizukawa}, Ryuki and {Odaka}, Hirokazu and {Sato}, Riki and {Tanimoto}, Atsushi and {Tashiro}, Makoto and {Yaqoob}, Tahir and {Yamada}, Satoshi},
        title = "{Unraveling the structure of the stratified ultra-fast outflows in PDS 456 with XRISM}",
      journal = {\pasj},
         year = 2025,
        month = jul,
          doi = {10.1093/pasj/psaf070},
archivePrefix = {arXiv},
       eprint = {2506.05273},
 primaryClass = {astro-ph.HE},
       adsurl = {https://ui.adsabs.harvard.edu/abs/2025PASJ..tmp...73X}
}

@ARTICLE{Gehrels1986,
       author = {{Gehrels}, N.},
        title = "{Confidence Limits for Small Numbers of Events in Astrophysical Data}",
      journal = {\apj},
         year = 1986,
        month = apr,
       volume = {303},
        pages = {336},
          doi = {10.1086/164079},
       adsurl = {https://ui.adsabs.harvard.edu/abs/1986ApJ...303..336G}
}

@ARTICLE{moravec22,
       author = {{Moravec}, Emily and {Svoboda}, Ji{\v{r}}{\'\i} and {Borkar}, Abhijeet and {Boorman}, Peter and {Kynoch}, Daniel and {Panessa}, Francesca and {Mingo}, Beatriz and {Guainazzi}, Matteo},
        title = "{Do radio active galactic nuclei reflect X-ray binary spectral states?}",
      journal = {\aap},
         year = 2022,
        month = jun,
       volume = {662},
          eid = {A28},
        pages = {A28},
          doi = {10.1051/0004-6361/202142870},
archivePrefix = {arXiv},
       eprint = {2202.11116},
 primaryClass = {astro-ph.GA},
       adsurl = {https://ui.adsabs.harvard.edu/abs/2022A&A...662A..28M}
}

@ARTICLE{Laurenti26,
       author = {{Laurenti}, M. and {Tombesi}, F. and {Cond{\`o}}, P. and {Gaspari}, M. and {Nicastro}, F. and {Torresi}, E. and {Luminari}, A. and {Piconcelli}, E. and {Zappacosta}, L. and {Fukumura}, K. and {Lanzuisi}, G. and {Serafinelli}, R. and {Dadina}, M. and {Cappi}, M. and {Middei}, R. and {Arevalo Gonzalez}, F. and {Di Salvo}, F.},
        title = "{A song of lines and winds: Tracing the signatures of AGN outflows in X-rays}",
      journal = {\aap},
         year = 2026,
        month = jan,
       volume = {705},
          eid = {A240},
        pages = {A240},
          doi = {10.1051/0004-6361/202556131},
archivePrefix = {arXiv},
       eprint = {2512.06077},
 primaryClass = {astro-ph.HE},
       adsurl = {https://ui.adsabs.harvard.edu/abs/2026A&A...705A.240L}
}

\onecolumn

\begin{appendix} 

\section{Sample properties} \label{tabelle proprietà}
\begin{table*}[h!]
\centering
\caption{List of the 33 sources analyzed in this work with their basic accretion properties.}
  \renewcommand{\arraystretch}{1.25}
\begin{tabular}{l c c c c c}
\hline
Source & z & $log(M_{BH})$ & $log(L_{Bol})$ & $\lambda_{\rm Edd}$ & H\\
\hline
(1) & (2) & (3) & (4) & (5) & (6)\\
\hline
\hline
3C 120 & 0.0330 & $7.74^{+0.04}_{-0.04}$(a) & $44.85^{+0.01}_{-0.01}$ & $0.10^{+0.01}_{-0.01}$ & $0.82^{+0.01}_{-0.01} $\\

3C 273 & 0.1583 & $8.84^{+0.08}_{-0.11}$(a) & $46.20^{+0.01}_{-0.02}$ & $0.18^{+0.05}_{-0.03}$ &  $0.79^{+0.01}_{-0.01}  $\\
   
3C 382 & 0.0556 & $8.98\pm0.47$(b) & $45.37^{+0.01}_{-0.04}$ & $0.02^{+0.04}_{-0.01}$&  $0.58^{+0.01}_{-0.01}  $\\
   
3C 390.3 & 0.0561 & $8.64^{+0.04}_{-0.05}$(a) & $45.19^{+0.01}_{-0.01}$ & $0.03^{+0.01}_{-0.01}$ &  $0.94^{+0.01}_{-0.01}  $\\
   
ARK 120 & 0.0327 & $8.07^{+0.05}_{-0.06}$(a) & $45.11^{+0.01}_{-0.01}$ & $0.09^{+0.01}_{-0.01}$ &  $0.35^{+0.01}_{-0.01}  $\\
   
ARK 564 & 0.0247  & $6.41^{+0.05}_{-0.04}$(k) & $44.29^{+0.01}_{-0.01}$ & $0.59^{+0.06}_{-0.07}$ &  $0.79^{+0.01}_{-0.1}  $ \\
   
ESO511-G030 & 0.0224& $7.23^{+0.05}_{-0.07}$(e) & $44.20^{+0.01}_{-0.01}$ & $0.07^{+0.01}_{-0.01}$ &  $0.61^{+0.01}_{-0.01}  $\\
  
Fairall 9 & 0.0461 & $8.30^{+0.08}_{-0.12}$(c) & $45.08^{+0.05}_{-0.03}$ & $0.05^{+0.02}_{-0.01}$ &  $0.40^{+0.01}_{-0.01}  $\\
   
HE 1029-1401 & 0.0858 & $8.85^{+0.05}_{-0.04}$(e) & $45.76^{+0.01}_{-0.05}$ & $0.06^{+0.01}_{-0.01}$ & $ 0.27^{+0.01}_{-0.01}   $\\
   
HE 1143-1810 & 0.0329 & $7.39\pm0.03$(e) & $44.76^{+0.01}_{-0.01}$ & $0.19^{+0.01}_{-0.01}$ &  $0.52^{+0.01}_{-0.01} $\\
   
IGRJ19378-0617 & 0.0102 & $6.40^{+0.17}_{-0.20}$(f) & $43.51^{+0.01}_{-0.01}$ & $0.10^{+0.06}_{-0.03}$ &  $0.69^{+0.01}_{-0.01}  $\\
  
Mkn 766 & 0.0129 & $6.82^{+0.05}_{-0.06}$(c) & $43.61^{+0.01}_{-0.01}$ & $0.05^{+0.01}_{-0.01}$ & $ >0.99 $\\

MR 2251-178 & 0.0640 & $8.19^{+0.08}_{-0.11}$(e) & $45.55^{+0.02}_{-0.01}$ & $0.19^{+0.05}_{-0.03}$ &  $0.65^{+0.01}_{-0.01}  $\\
   
Mrk 110 & 0.0353 & $7.29^{+0.10}_{-0.10}$(a) & $44.75^{+0.01}_{-0.01}$ & $0.23^{+0.06}_{-0.05}$ &  $0.61^{+0.01}_{-0.01}  $\\
   
Mrk 279 & 0.0304 & $7.43^{+0.10}_{-0.13}$(a) & $44.65^{+0.01}_{-0.01}$ & $0.13^{+0.05}_{-0.03}$ &  $0.55^{+0.01}_{-0.01}  $\\
   
Mrk 335 & 0.0258 & $7.23^{+0.04}_{-0.04}$(c) & $45.17^{+0.02}_{-0.01}$ & $1.02^{+0.07}_{-0.06}$ &  $0.18^{+0.01}_{-0.01} $ \\
   
Mrk 509 & 0.0344 & $8.05\pm0.03$(a) & $44.97^{+0.01}_{- 0.01}$ & $0.07^{+0.01}_{-0.01}$ &  $0.18^{+0.01}_{-0.01}  $\\
   
Mrk 841 & 0.0364 & $8.16^{+0.05}_{-0.06}$(e) & $44.06^{+0.01}_{-0.03}$ &  $0.01^{+0.01}_{-0.01}$ &  $0.34^{+0.01}_{-0.01}  $\\
   
Mrk 926 & 0.0470 & $7.98\pm0.06$(e) & $45.16^{+0.05}_{-0.02}$ & $0.12^{+0.02}_{-0.02}$ &  $0.86^{+0.02}_{-0.01}  $\\
   
Mrk 1044 & 0.0164 & $6.45^{+0.12}_{-0.13}$(j) & $44.07^{+0.06}_{-0.03}$ & $0.33^{+0.13}_{-0.08}$ &  $0.48^{+0.02}_{-0.01} $ \\
   
NGC 985 & 0.0431 & $8.11^{+0.03}_{-0.01}$(e) & $44.94^{+0.01}_{-0.01}$ & $0.05^{+0.01}_{-0.01}$ &  $0.39^{+0.01}_{-0.01}   $\\
   
NGC 1566 & 0.0050 & $6.92\pm0.08$(a) & $43.48^{+0.04}_{-0.01}$ & $0.03^{+0.01}_{-0.01}$ &  $0.71^{+0.02}_{-0.01}  $\\
   
NGC 2617 & 0.0142 & $7.38^{+0.09}_{-0.05}$(l) & $44.05^{+0.01}_{-0.01}$ & $0.04^{+0.01}_{-0.01}$ &  $0.85^{+0.01}_{-0.01}  $\\
   
NGC 4051 & 0.0024 & $6.13^{+0.08}_{-0.145}$(a) & $41.96^{+0.01}_{-0.01}$ & $0.01^{+0.06}_{-0.10}$ &  $>0.99$\\
   
NGC 4593 & 0.0083 & $6.88^{+0.07}_{-0.07}$(a) & $43.62^{+0.02}_{-0.01}$ & $0.04^{+0.01}_{-0.01}$ & $ 0.84^{+0.01}_{-0.01}  $\\
  
NGC 5548 & 0.0172 & $7.69\pm0.02$(c) & $45.12^{+0.01}_{-0.01}$ & $0.21^{+0.01}_{-0.01}$ &  $0.32^{+0.01}_{-0.01}  $\\
   
NGC 7213 & 0.0058 & $7.13^{+0.02}_{-0.03}$(e) & $42.91^{+0.09}_{-0.01}$ & $0.01^{+0.01}_{-0.01}$ &  $>0.99$\\
   
NGC 7469 & 0.0163 & $6.96^{+0.05}_{-0.05}$(a) & $44.26^{+0.01}_{-0.01}$ & $0.16^{+0.02}_{-0.02}$ & $ 0.51^{+0.01}_{-0.01}  $\\
   
NGC 7603 & 0.0288 & $8.22^{+0.02}_{-0.03}$(i) & $44.82^{+0.01}_{-0.03}$ & $0.03^{+0.01}_{-0.01}$ &  $0.24^{+0.01}_{-0.01} $\\
   
PG 2304+042 & 0.0430 & $7.58^{+0.02}_{-0.01}$(i) & $44.63^{+0.01}_{-0.05}$ & $0.09^{+0.01}_{-0.01}$ &  $0.71^{+0.01}_{-0.02}  $ \\
  
PKS0558-504 & 0.1372 & $7.62^{+0.26}_{-0.33}$(e) & $45.83^{+0.03}_{-0.01}$ & $1.29^{+1.47}_{-0.58}$ & $ 0.49^{+0.01}_{-0.01}  $\\
   
SBS 1301+540  & 0.0301 & $7.67^{+0.09}_{-0.02}$(e) & $44.23^{+0.02}_{-0.04}$ & $0.03^{+0.02}_{-0.01}$ & $ 0.91^{+0.01}_{-0.02}   $\\
   
UGC 3973 & 0.0222 & $7.61^{+0.11}_{-0.14}$(a) & $44.21^{+0.07}_{-0.04}$ & $0.03^{+0.01}_{-0.01}$ & $ 0.71^{+0.01}_{-0.01}$\\
   \hline
\end{tabular}
\vspace{2mm}
\label{Table 4.1}
\tablefoot{(1) Source name; (2) redshift; (3) log of the black hole masses in $\Msun$; (4) log of $L_{Bol}$ in erg/s; (5) $\lambda_{Edd}$; (6) $H=\frac{L_P}{L_P + L_{D}}$, all from \citet{Svoboda17}.
SMBH masses estimated according to: reverberation mapping method ((a) \cite{Bentz18}, (b) \cite{Fausnaugh17}, (c) \cite{Bentz15}, (f) \cite{Robinson21}, (i) \cite{zastrocky2024}, (j) \cite{Du15}, (k) \cite{Botte2004}, (l) \cite{Feng2021}) or single epoch $H\beta$-based
((d) \cite{Vestergaard06}, (e) \cite{Mej22}).
}
\end{table*}

\begin{table*}[h!]
\centering
\caption{List of the observations and X-ray properties. }
  \renewcommand{\arraystretch}{1.25}
\begin{tabular}{l c c c c c}
\hline
Source & Obs ID & Duration ({Net $t_{exp}$}) & Net Counts & $L_{2-12 \rm keV}$ & $F_{2-12 \rm keV}$\\
\hline
(1) & (2) & (3) & (4) & (5) & (6)\\
\hline
\hline
3C 120 & 0152840101 & 13.38 (8.66) & 420903 & $44.11^{+0.01}_{-0.01}$  & $5.17^{+0.01}_{-0.03}$ \\ 
  
3C 273 & 0136550101 & 8.98 (6.18) & 594694 & $45.87^{+0.01}_{-0.01}$  & $11.29^{+0.05}_{-0.31}$ \\ 
   
3C 382 & 0506120101 & 3.94 (2.22) & 94529 & $44.51^{+0.01}_{-0.02}$ & $4.49^{+0.01}_{-0.19}$ \\ 
   
3C 390.3 & 0203720201 & 7.04 (3.60) & 141231 & $44.52^{+0.01}_{-0.01}$ & $4.46^{+0.02}_{-0.07}$ \\ 
   
ARK 120 & 0721600401 & 13.33 (8.53) & 385897 & $44.04^{+0.01}_{-0.01}$   & $4.47^{+0.03}_{-0.01}$ \\ 
   
ARK 564* & 0830540101 & 11.49 (7.62) & 125977 & $43.31^{+0.04}_{-0.01}$ & $1.45^{+0.02}_{-0.03}$ \\ 
   
ESO511-G030 & 0502090201 & 11.23 (7.61) & 163506 & $43.41^{+0.01}_{-0.01}$ & $2.28^{+0.01}_{-0.01}$ \\ 
   
Fairall 9 & 0741330101& 14.14 (8.75) & 201718 &  $44.10^{+0.02}_{-0.01}$   & $2.44^{+0.17}_{-0.01}$ \\ 
   
HE 1029-1401* & 0890410101 & 10.39 (7.14) & 78847 & $43.47^{+0.01}_{-0.05}$   & $1.99^{+0.07}_{-0.06}$ \\ 
   
HE 1143-1810 & 0201130201 & 3.41 (2.17) & 67419 & $43.92^{+0.01}_{-0.01}$   &$ 3.33^{+0.01}_{-0.02} $\\ 
   
IGRJ19378-0617* & 0761870201 & 14.14 (9.32) & 223573 &  $42.76^{+0.01}_{-0.02}$  & $2.44^{+0.01}_{-0.02}$ \\ 
   
Mkn 766 & 0109141301& 12.99 (8.75) & 230483 & $42.98^{+0.01}_{-0.01}$   & $2.59^{+0.03}_{-0.01}$ \\ 
   
MR 2251-178 & 0670120201 & 13.37 (9.15) & 384517 & $44.67^{+0.02}_{-0.01}$   & $4.87^{+0.03}_{-0.01}$ \\ 
   
Mrk 110 & 0201130501 & 4.74 (3.28) & 102229 & $43.98^{+0.01}_{-0.01}$& $3.36^{+0.02}_{-0.02}$  \\ 
   
Mrk 279 & 0302480401 & 5.98 (3.87) & 108012 & $43.82^{+0.01}_{-0.01}$   & $3.10^{+0.01}_{-0.02}$ \\ 
  
Mrk 335* & 0306870101 & 13.32 (8.91) & 177893 & $43.49^{+0.02}_{-0.01}$  & $ 2.02^{+0.06}_{-0.09}$ \\ 

Mrk 509 & 0306090201 & 8.59 (5.97) & 236968 & $44.06^{+0.01}_{-0.01}$   & $4.23^{+0.01}_{-0.26}$ \\ 
   
Mrk 841*& 0882130301 & 13.20 (8.41) & 127824 & $43.69^{+0.01}_{-0.03}$  & $1.60^{+0.05}_{-0.03} $\\ 
   
Mrk 926 & 0790640101 & 5.90 (3.41) & 168691 & $44.47^{+0.02}_{-0.01}$  & $5.8^{+0.1}_{-0.3} $\\ 
 
Mrk 1044 & 0824080501 & 14.14 (9.28) & 152148 & $43.01^{+0.02}_{-0.01}$   & $1.69^{+0.02}_{-0.05}$ \\ 
  
NGC 985 & 0743830601 & 12.20 (8.91) & 135291  & $43.90^{+0.01}_{-0.01}$   & $1.83^{+0.01}_{-0.01}$ \\ 
   
NGC 1566 & 0820530401 & 10.80 (7.40) & 119130 & $42.54^{+0.01}_{-0.01}$   &  $6.3^{+0.2}_{-0.1}$ \\ 
   
NGC 2617 & 0701981901 & 3.46 (1.47) & 72615 & $43.40^{+0.01}_{-0.01}$   & $5.57^{+0.02}_{-0.20}$  \\ 
  
NGC 4051 & 0109141401 & 12.20 (7.08) & 177329 & $41.51^{+0.01}_{-0.01} $  &$ 2.63^{+0.01}_{-0.01}$\\
  
NGC 4593* & 0784740101 & 14.21 (9.84) & 233106 & $42.60^{+0.02}_{-0.01}$   &  $2.62^{+0.02}_{-0.01} $\\ 
   
NGC 5548 & 0089960301 & 9.58 (5.71) & 223736 & $43.47^{+0.01}_{-0.01}$  &$ 4.48^{+0.01}_{-0.03} $\\ 
   
NGC 7213* & 0605800301 & 13.25 (8.73) & 106218 & $42.02^{+0.09}_{-0.01}$   &$1.41^{+0.01}_{-0.01} $ \\ 
   
NGC 7469 & 0760350801 & 10.16 (6.50) & 219175 & $43.33^{+0.01}_{-0.01}$   & $3.59^{+0.04}_{-0.01} $ \\
   
NGC 7603 & 0305600601 & 1.68 (1.14) & 21710  & $43.61^{+0.01}_{-0.01}$  &$ 2.16^{+0.02}_{-0.01} $ \\ 
  
PG 2304+042 & 0783272701 & 1.30 (0.86) & 12548  & $43.86^{+0.01}_{-0.02}$  &  $1.71^{+0.02}_{-0.01}$ \\
 
PKS0558-504 & 0555170401 & 12.92 (8.61) & 139660 & $44.91^{+0.02}_{-0.01}$  & $1.58^{+0.01}_{-0.01}$ \\ 
   
SBS 1301+540 & 0312192001 & 1.19 (0.75) & 12430 & $43.57^{+0.01}_{-0.02}$   &  $1.80^{+0.04}_{-0.08}$ \\ 
  
UGC 3973* & 0502091001 & 8.85 (5.29) & 36227 & $43.03^{+0.07}_{-0.03}$   &$0.97^{+0.02}_{-0.01}$\\
   \hline
\end{tabular}
\vspace{2mm}
\tablefoot{Column 1: Source name. Column 2: \xmm observational ID. Column 3: total and net exposure time for the pn camera after removing soft proton flare-dominated intervals, 
in units of $10^4$ seconds. Column 4: number of counts in the pn 2-12 keV band. Column 5: log of the luminosity in $10^{43}$ erg/s in the 2-12 keV band. Column 6: flux in $10^{-11}$ (erg/cm$^{2}$/s) in the 2-12 keV band.
* A more recent OBSID, with respect to \citet{Svoboda17} was used, given the significantly longer exposure time.
}
\label{Table 4.2}
\end{table*}

\newpage

\section{Continuum modeling} \label{figure modelli}
In this section, we present the best-fit models used to reproduce AGN emission in the spectra of sources with UFOs. The BM (described in Sect \ref{Continuum modeling}) is made of a combination of emission components due to direct continuum and reflection. Some of the sources needed additional components in order to achieve a good description of the spectra: 

\begin{itemize}
    \item in Mkn 766 (\ref{mkn766_mrk110}, left), we added a Gaussian emission line to model the broad component from $K{\alpha}$, taking into account relativistic effects;
    \item MRK 110 (\ref{mkn766_mrk110}, right), we added a Gaussian emission line to model the broad component from $K{\alpha}$, taking into account relativistic effects, a second power-law to reproduce the soft-excess; 
    \item NGC 2617 (\ref{ngc2617_ngc5548}, left), we added a second power-law to reproduce the soft-excess;
    \item NGC 5548 (\ref{ngc2617_ngc5548}, right), the BM was sufficient to model the spectrum;
    \item NGC 7469 (\ref{ngc7469_pg2304}), we added a Gaussian emission line to model the broad component from $K{\alpha}$, taking into account relativistic effects, a second power-law to reproduce the soft-excess; 
    \item PG 2304 (\ref{ngc7469_pg2304}), we added a Gaussian emission line to model the broad component from $K{\alpha}$, taking into account relativistic effects.
\end{itemize}

\begin{figure*}[h]
    \centering
    \includegraphics[scale=1, angle=0, width=8cm,height=6cm,keepaspectratio]{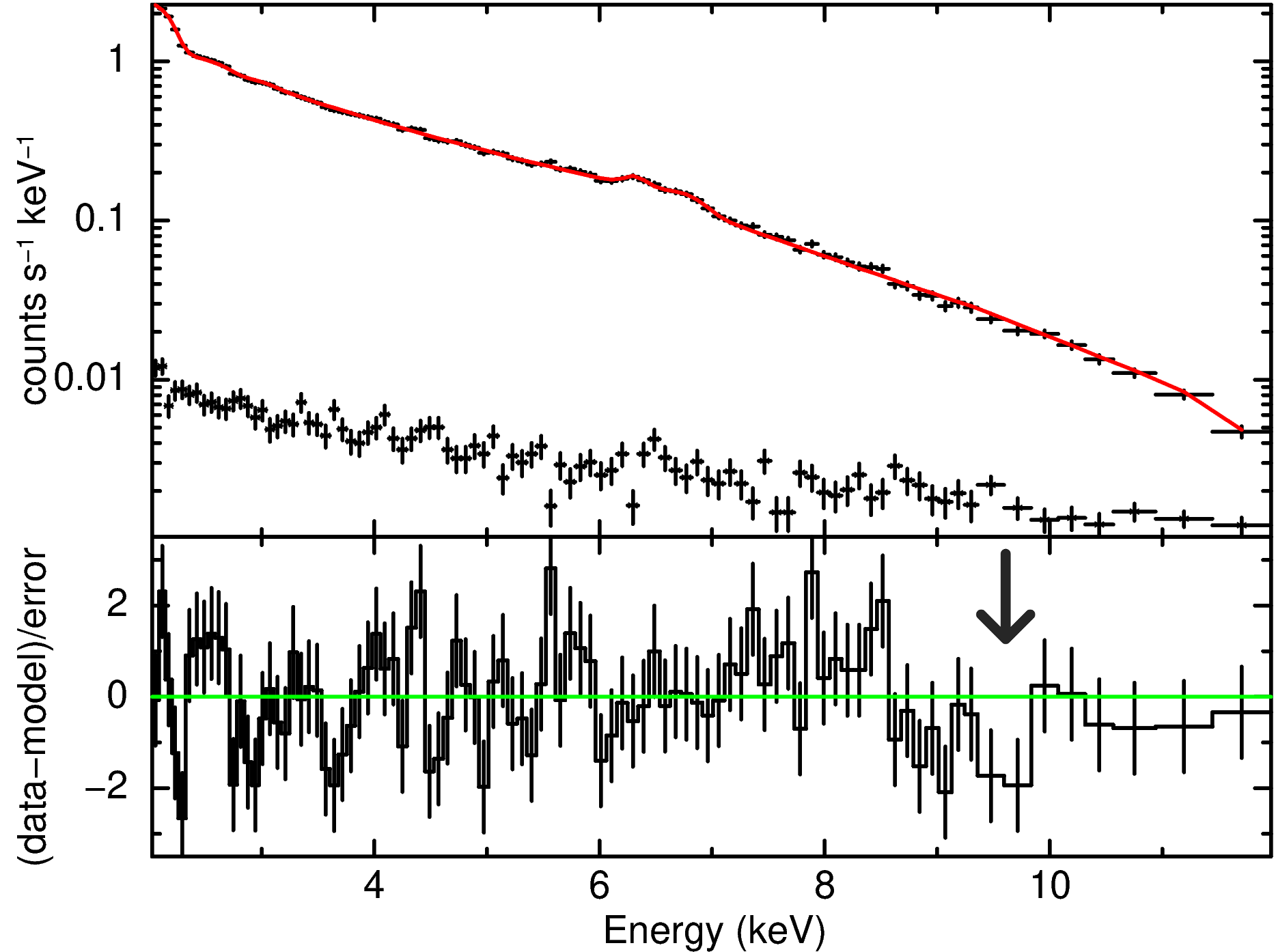}
    \hspace{0.5cm}
    \includegraphics[scale=1, angle=0, width=8cm,height=6cm,keepaspectratio]{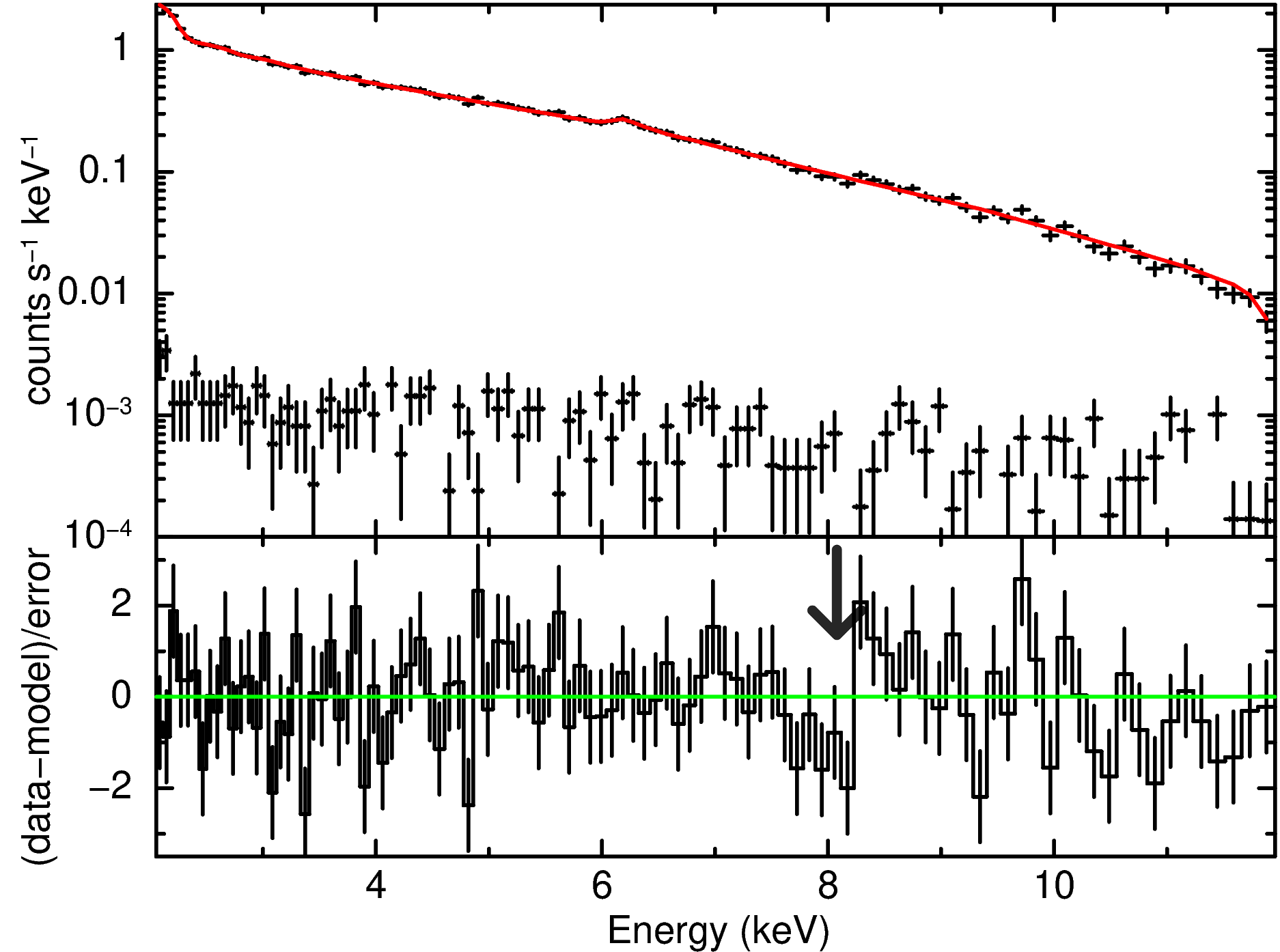}
    \caption{Left: Best-fit Baseline Model applied to Mkn~766 data (top) and residuals (bottom). 
    Right: Best-fit Baseline Model applied to  Mrk~110 data (top) and residuals (bottom). 
    The UFO's absorption features are highlighted with an arrow.}
    \label{mkn766_mrk110}
\end{figure*}

\begin{figure*}[h]
    \centering
    \includegraphics[scale=1, angle=0, width=8cm,height=6cm,keepaspectratio]{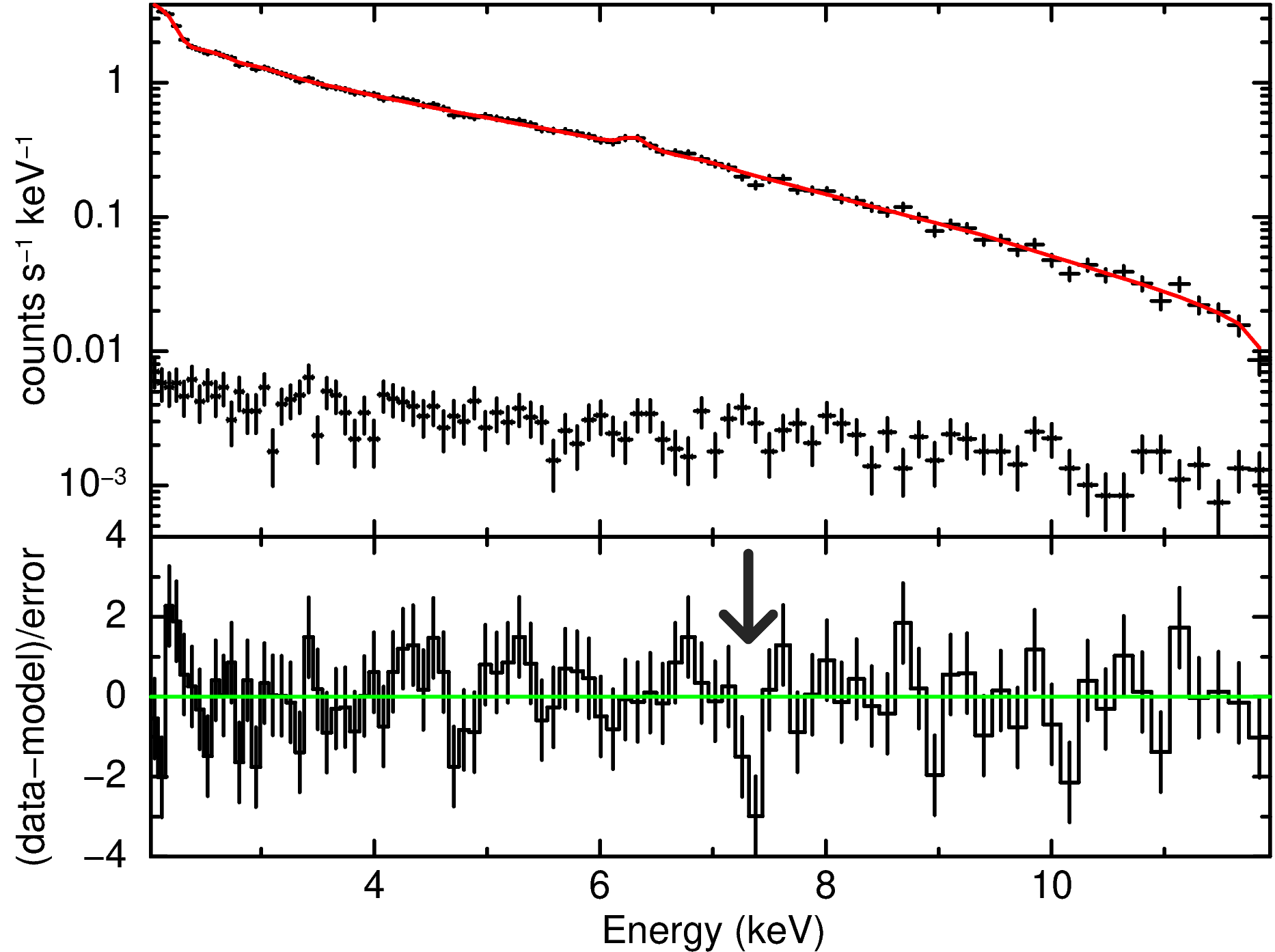}
    \hspace{0.5cm}
    \includegraphics[scale=1, angle=0, width=8cm,height=6cm,keepaspectratio]{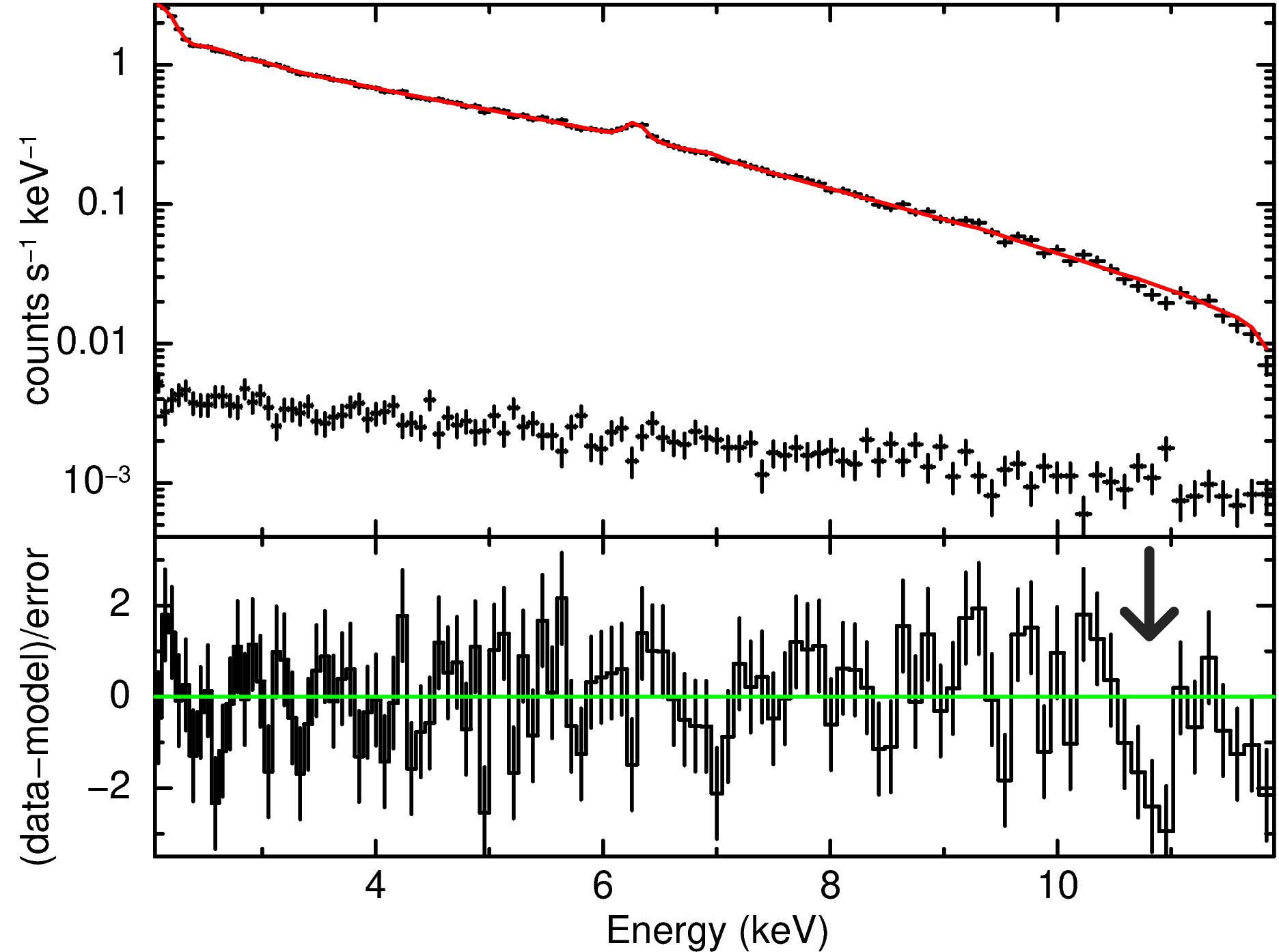}
    \caption{Left: Best-fit Baseline Model applied to NGC~2617 data (top) and residuals (bottom). 
    Right: Best-fit Baseline Model applied to NGC~5548 data (top) and residuals (bottom). 
    The UFO's absorption features are highlighted with an arrow.}
    \label{ngc2617_ngc5548}
\end{figure*}

\begin{figure*}[h]
    \centering
    \includegraphics[scale=1, angle=0, width=8cm,height=6cm,keepaspectratio]{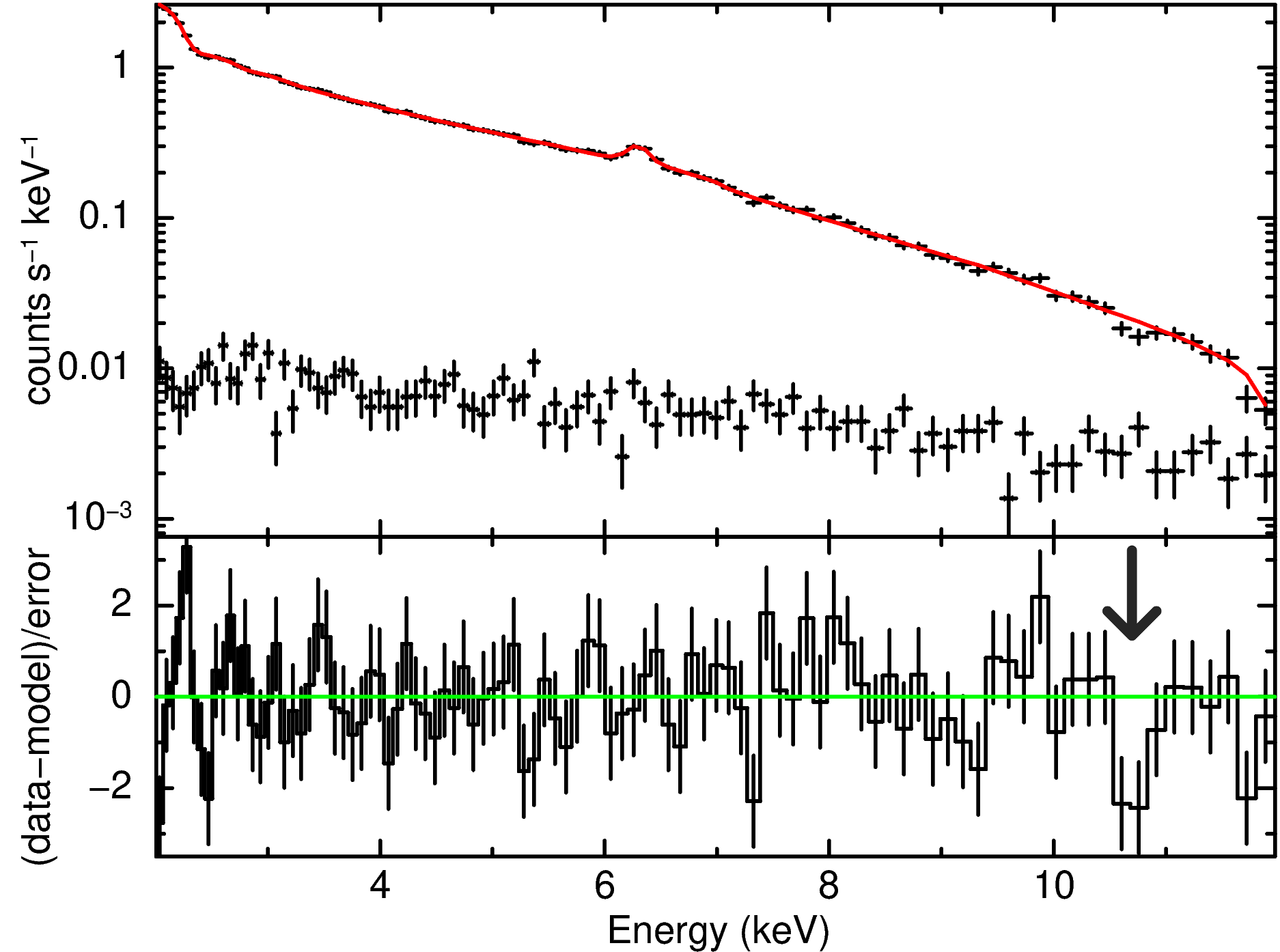}
    \hspace{0.5cm}
    \includegraphics[scale=1, angle=0, width=8cm,height=6cm,keepaspectratio]{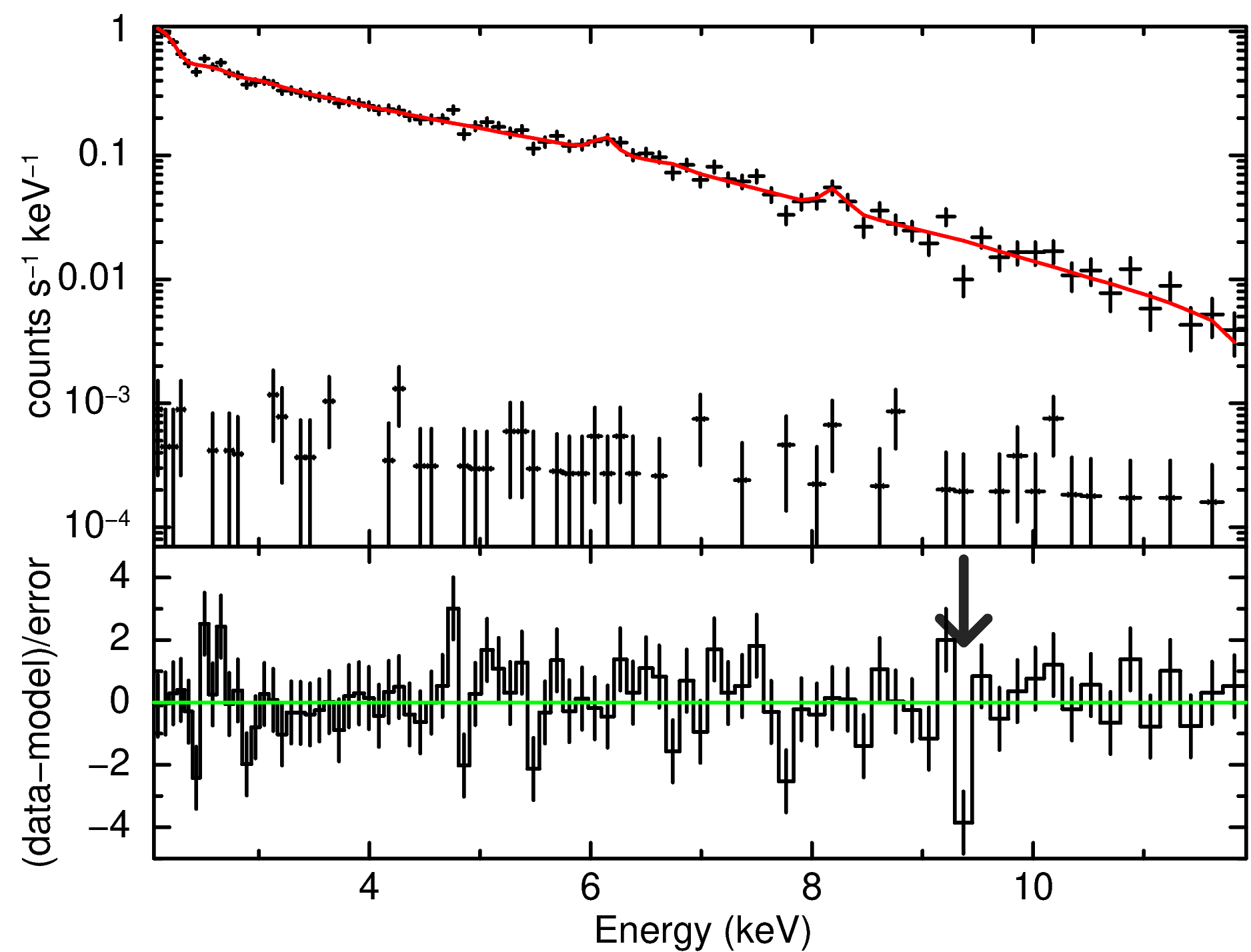} 
    \caption{Left: Best-fit Baseline Model applied to NGC~7469 data (top) and residuals (bottom). 
    Right: Best-fit Baseline Model applied to PG~2304 data (top) and residuals (bottom). 
    The UFO's absorption features are highlighted with an arrow.}
    \label{ngc7469_pg2304}
\end{figure*}

\newpage
\section{Comparison to BXA} 
\label{bxa}

In this section, we compare the determination of the best-fit parameters described in Sect.~\ref{sect:abs} with the Bayesian X-ray Analysis (BXA; \cite{buchner16}). BXA employs the nested sampling algorithm UltraNest for Bayesian parameter estimation. We have used Gaussian prior distributions for line energies, width, and depth, with distributions equal to the 99.7\% confidence error on the formal forward-folding best fit. As shown in Fig.~\ref{fig:conf}, the measurements with the two methods are consistent within the 90\% level statistical uncertainties.

\begin{figure}[ht]
    \centering
    \includegraphics[scale=1, angle=0, width=6.0cm,height=9cm,keepaspectratio]{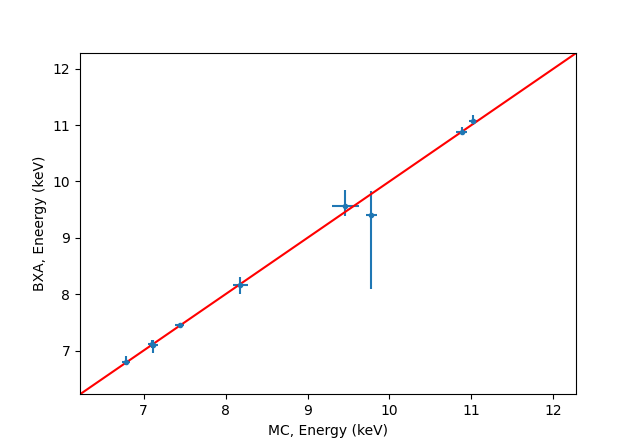}
    \includegraphics[scale=1, angle=0, width=6.0cm,height=9cm,keepaspectratio]{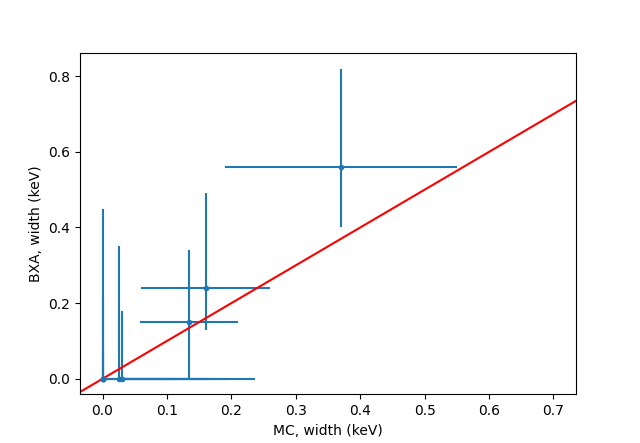}
    \includegraphics[scale=1, angle=0, width=6.0cm,height=9cm,keepaspectratio]{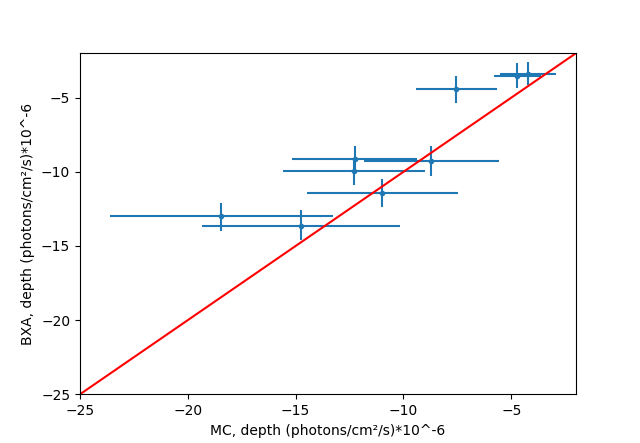}
    \caption{Comparison between the determination of the absorption line spectral parameters using a standard forward-folding approach and BXA. The red line is the bisector.}
    \label{fig:conf}
\end{figure}

\section{\nustar validation of UFO detections}
\label{app:Nustar}

\begin{figure*}[t]
    \centering
    \includegraphics[scale=1, angle=0, width=13cm,height=6cm,keepaspectratio]{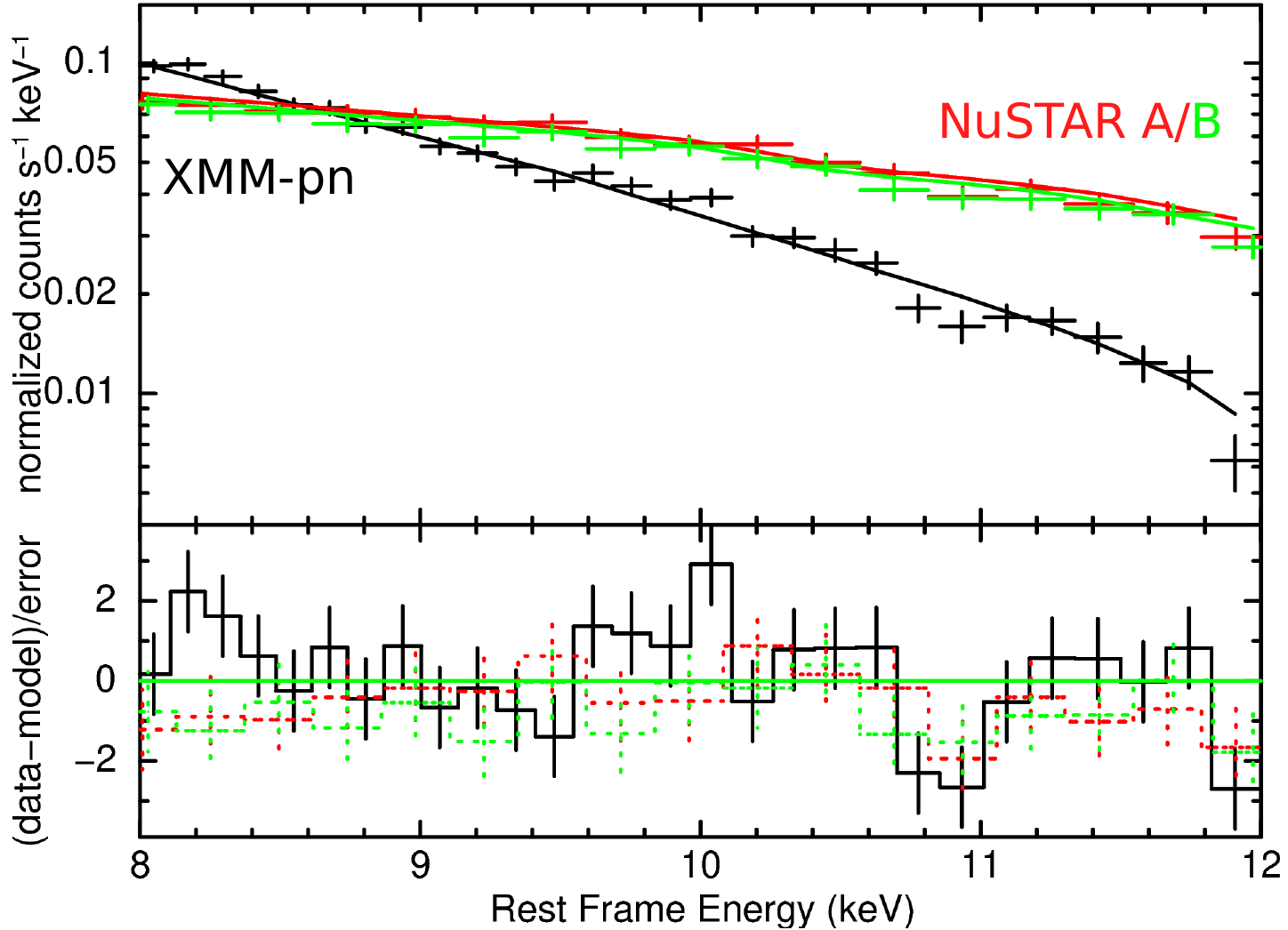}
    \hspace{0.5cm}
    \includegraphics[scale=1, angle=0, width=13cm,height=6cm,keepaspectratio]{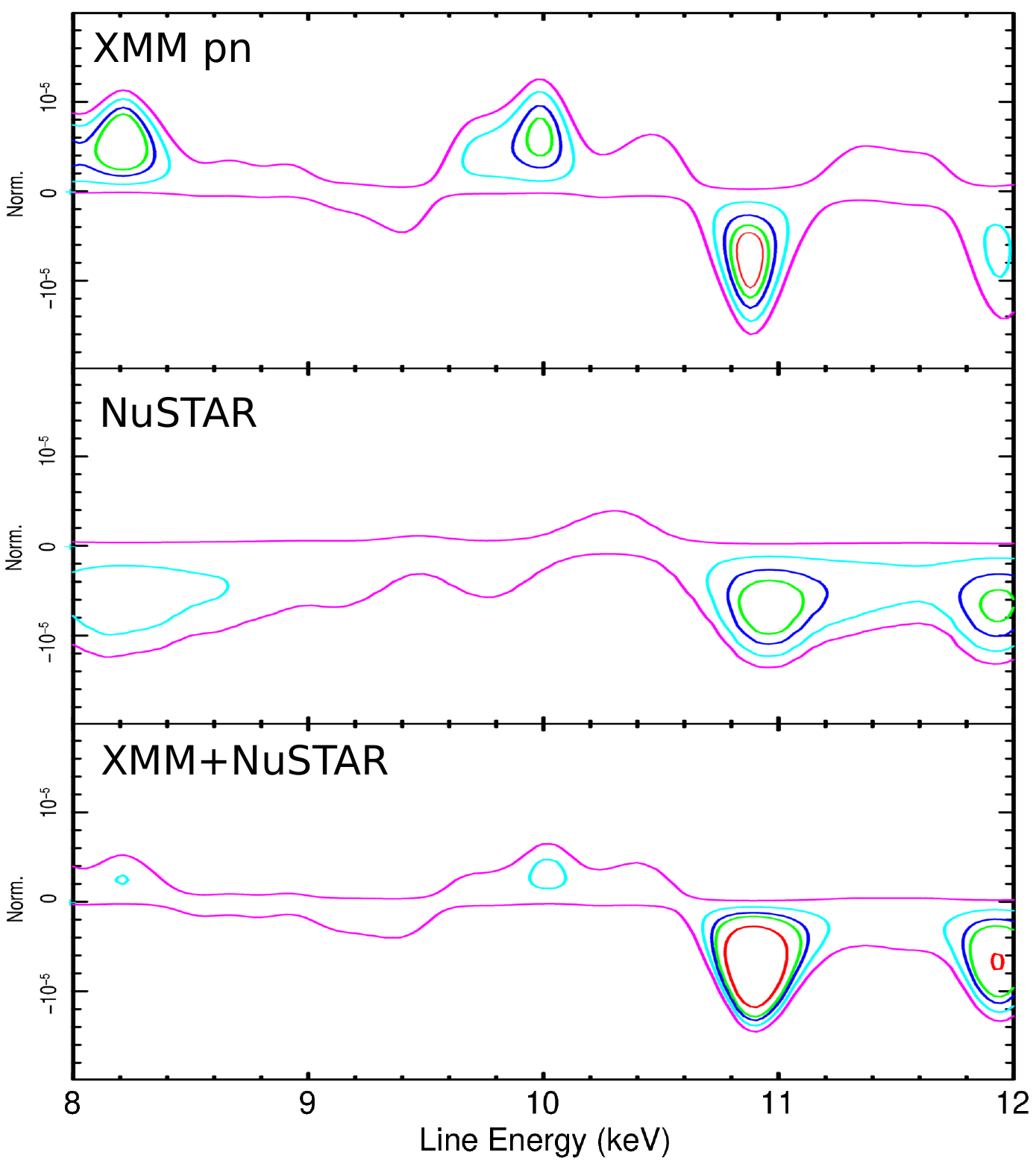} 
    \caption{\xmm pn and \nustar FPMA-B spectra (left) and line scan (right) of NGC~7469 in the 8-12 keV band for Epoch 1, alongside the best-fit Baseline Model and corresponding residuals. The high-velocity UFO absorption features are clearly visible in both instruments at $\sim10.8$ keV.}
    \label{epoch1}
\end{figure*}

\begin{figure*}[t]
    \centering
    \includegraphics[scale=1, angle=0, width=13cm,height=6cm,keepaspectratio]{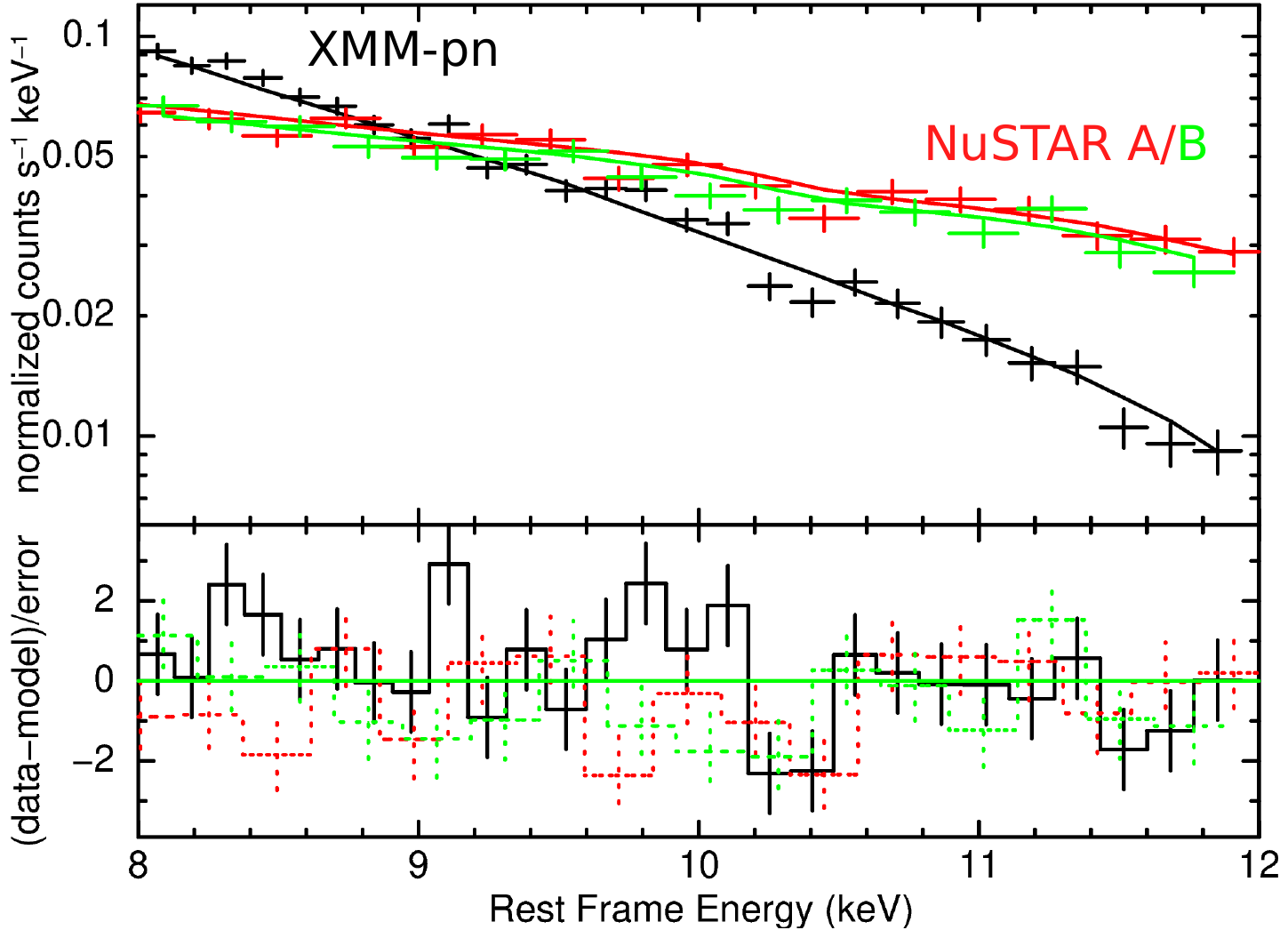}
    \hspace{0.5cm}
    \includegraphics[scale=1, angle=0, width=13cm,height=6cm,keepaspectratio]{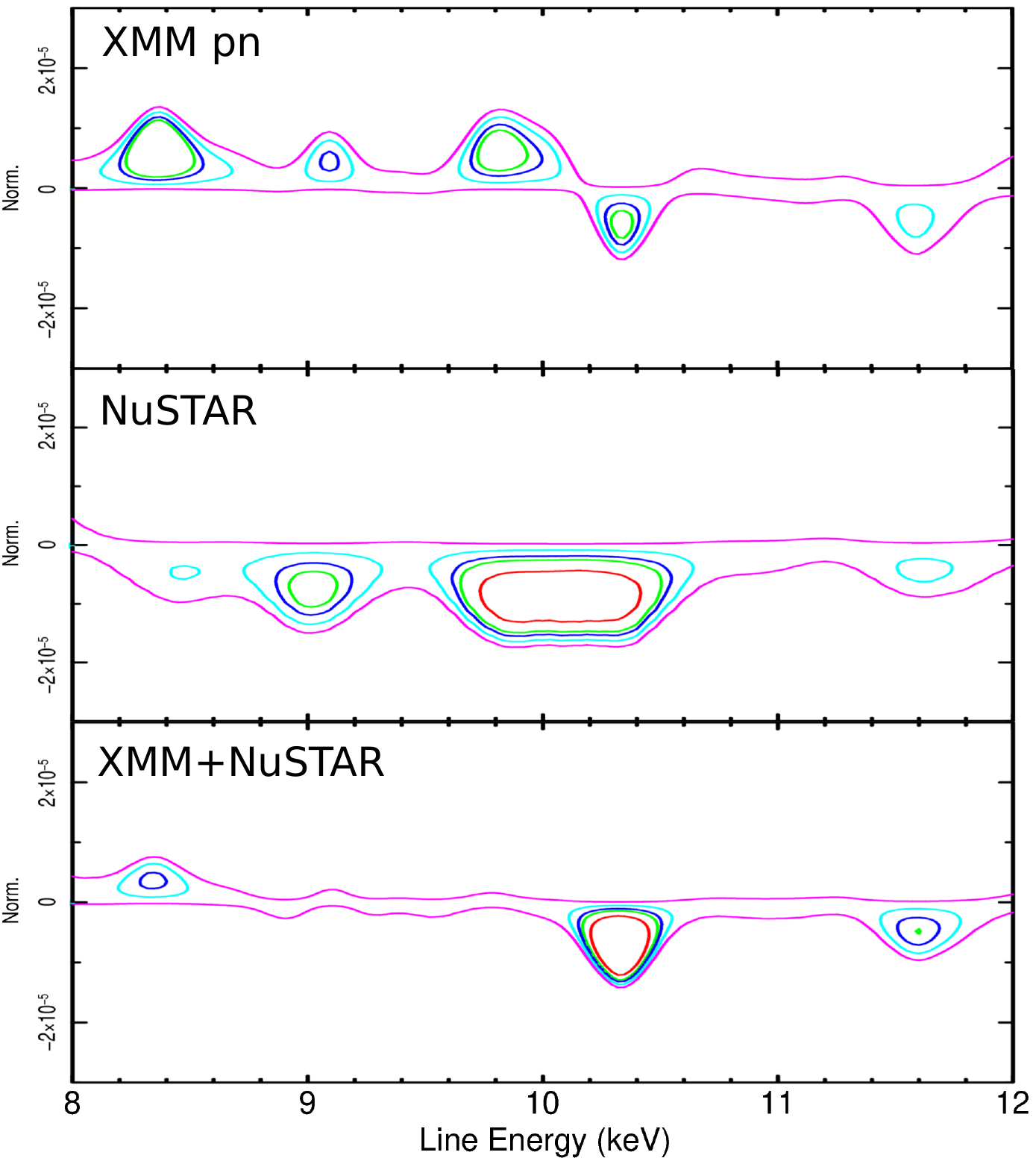} 
    \caption{\xmm pn and \nustar FPMA-B spectra (left) and line scan (right) of NGC~7469 in the 8-12 keV band for Epoch 2, alongside the best-fit Baseline Model and corresponding residuals. The high-velocity UFO absorption features are clearly visible in both instruments at $\sim10.3$ keV.}
    \label{epoch2}
\end{figure*}

As discussed in \Cref{Sec. 5}, the effective area and calibration of the \xmm EPIC-pn camera become increasingly uncertain above 10 keV. To validate the robustness of our highest-energy UFO detections against potential instrumental artifacts, we cross-checked our results using simultaneous \nustar observations, which offer superior sensitivity and well-constrained calibration in the hard X-ray band, at the expense of a factor 3-4 worse energy resolution.

We verified the availability of strictly simultaneous \nustar data for all sources in our sample exhibiting UFOs in the 9-12 keV band (Mkn~766, NGC~5548, NGC~7469, and PG~2304+042). Among these, only NGC~7469 possesses contemporary \xmm and \nustar observations, obtained during a coordinated campaign in 2015. While a comprehensive analysis of all available coordinated pointings from that campaign and the derivation of detailed variability constraints are beyond the scope of this paper, we use them here as a crucial diagnostic test for our $>10$ keV detection methodology.

For this validation, we selected two specific epochs from the 2015 campaign: Epoch 1, the \nustar observation (OBS-ID 60101001014) strictly simultaneous with the primary \xmm observation analyzed in the main text (OBSID 0760350801).
Epoch 2, an additional set of simultaneous pointings taken between December 22-24 (\xmm OBSID 0760350601 and \nustar OBSID 60101001008). In this second epoch, the fast UFO feature is similarly prominent, yielding an improvement in the Cash statistic of $\Delta C=11.8$ in the \xmm data alone.

To assess the consistency of the absorption feature across both observatories, we restricted our spectral modeling to the $7-12$ keV band. We applied the BM to both datasets and extracted the corresponding residuals. Furthermore, we performed a systematic line scan across this energy range by stepping a narrow ($\sigma=10 \rm eV$) Gaussian absorption line through the data. This scan was executed three times for each epoch: on the \xmm data alone, on the \nustar data alone, and on the combined \xmm and \nustar spectra.

The resulting best-fit spectra, residuals, and $\Delta C$ contour scans for both epochs are presented in \Cref{epoch1} and \Cref{epoch2}. The presence of these highly blueshifted absorption features is confirmed in the independent \nustar data, with the lines detected at energies of $\sim10.8$ keV in Epoch 1 and $\sim10.3$ keV in Epoch 2. For each epoch, the line energies and depths are perfectly consistent between the \xmm and \nustar observations. This cross-instrument agreement clearly demonstrates that the $>10$ keV UFO detections identified in the EPIC-pn data are robust astrophysical features and are not driven by instrumental noise or by effective-area calibration drops. We acknowledge that the \nustar cross-validation presented here is limited to a single source, as it is the only target in our sample with strictly simultaneous hard X-ray coverage in the $>10$ keV band where our new UFO detections lie. While this successfully demonstrates that our methodology does not produce spurious detections from EPIC-pn calibration artifacts, a comprehensive multi-instrument confirmation of all four $\vout>0.3\rm c$ UFOs will require dedicated follow-up, which we plan to pursue with both archival \nustar data (where available, albeit nonsimultaneous) and upcoming XRISM/Resolve observations. 

\newpage

\section{Discarded absorption lines}
As discussed in Sect. \ref{Statistical analysis and robustness of results}, we considered an absorption line significantly detected if its addition in the model led to a statistical improvement from the continuum model of $\Delta C_{stat}>7.815$, corresponding to a significance of 95\% for a $\chi^2$ distribution with 3 degrees of freedom \citep{Lampton76}. Table \ref{discarded} reports possible UFO detections that were discarded as the absorption lines were detected with significance below this threshold (i.e., $\Delta C_{stat}<7.815$).
 
\begin{table}[h!]
\centering
\caption{Discarded absorption lines.}
  \renewcommand{\arraystretch}{1.25}
\begin{tabular}{c c c}
\hline
Source & $\Delta C_{stat}$ & $E_{out}$\\
\hline
(1) & (2) & (3)\\
\hline
\hline
3C273 & 6.64 &  $7.38\pm0.13$\\
   
Ark 120 & 5.77 & $10.80\pm0.60$\\
   
Fairall 9 & 5.71 & $9.96\pm0.11$\\
   
HE 1029-1401 & 4.61 & $11.08\pm0.30$ \\

IGRJ19378-0617 & 6.34 & $7.93\pm0.11$\\
   
Mrk 509 & 6.61 & $7.44\pm0.12$\\

NGC 4593 & 4.79 & $10.51\pm0.67$\\
   
NGC 7213 & 5.65 & $10.81\pm0.34$\\
   \hline
\end{tabular}
\vspace{2mm}
\tablefoot{(1) source name; (2) $\Delta C_{stat}$; (3) best-fit energy of absorption line, in keV.}
\label{discarded}
\end{table}

We decided to include in this section an absorption feature detected with high confidence (>98\%) in OBS:0830540101 of ARK 564. The properties of the UFO are shown in Table \ref{Table ark 564}. While the feature is detected with high statistics, the proper modeling goes beyond the simplistic models used in this analysis due to the known Fe K relativistic emission line observed in this source \citep{Kara17}.
Therefore, we delay the modeling of the detected absorption line in a forthcoming paper.

\begin{table*}[h!]
\centering
\caption{Properties of the absorption feature detected in ARK 564. }
  \renewcommand{\arraystretch}{1.25}
\begin{tabular}{c c c c c c c c c}
\hline
Source & $\Gamma$ & Energy & Width & |Depth| &  $\Delta$ Cstat & Significance & |EQW| & v/c \\
\hline
(1) & (2) & (3) & (4) & (5) & (6) & (7) & (8) & (9)\\
\hline
\hline
Ark 564 & $2.39_{-0.03}^{+0.03}$ & $7.35\pm0.10$ & $\le0.155$ & $3.95\pm1.95$ & 9.82 & 98.60\% & $32.3_{-7.2}^{+7.0}$  & $0.053\pm0.007$\\

   \hline
\end{tabular}
\vspace{2mm}
\tablefoot{Column 1: Source name. Column 2: rest-frame energy in keV. Column 3: width in keV. Column 4: depth (Gaussian normalization) in units of photons cm$^{-2}$s$^{-1}\cdot10^{-6}$. Column 5: $\Delta C_{stat}$. Column 6: significance from MC simulations (see Sect. \ref{Statistical analysis and robustness of results}). Column 7: EQW in eV. Column 8: velocity in units of $\rm c$. Uncertainties are at 95\%. EQW uncertainties at 68\%. }
\label{Table ark 564}
\end{table*}

\newpage

\section{Comparison with literature results}
Table \ref{Table 6.1} reports the comparison between our results with the outcomes of previous analysis on the same observations in our sample.

\begin{table}[h!]
\centering
\caption{Comparison with previous results on the same observations. }
  \renewcommand{\arraystretch}{1.15}
\begin{tabular}{c c c c}
\hline
Source &This work (v/c) & Previous works& Detection (v/c)\\
\hline
(1) & (2) & (3) & (4)\\
\hline
\hline
3C 120 & NO & \cite{Tombesi14} & NO \\
   
3C 390.3 & NO & \cite{Tombesi14} & NO \\
   
Ark 564 & NO & \cite{Tombesi10} & NO \\
   
ESO511-G030 & NO & \cite{Tombesi10} & NO\\
   
Mkn 766{\bf *} & $0.30\pm0.02$ & \cite{Tombesi10} & $0.091\pm0.004$\\
   
MRK 0110 & $0.16\pm0.01$ & \cite{Tombesi10} & NO\\
   
Mrk 335 & NO & \cite{Tombesi10} & NO \\
   
Mrk 509\textdagger & NO & \cite{Tombesi10} & $0.141\pm0.002$\\
   
Mrk 1044$\ddagger$ & NO & \cite{Xu2023} & NO \\
   
NGC 1566 & NO & \cite{jana21} & NO\\
   
NGC 2617 & $0.065\pm0.008$ & \cite{Giustini17} & redshifted\\
   
NGC 4051 & $0.018\pm0.006$ & \cite{Tombesi10} & $0.018\pm0.004$\\
   
NGC 4593 & NO & \cite{Tombesi10} & NO\\
   
NGC 5548 & $0.021\pm0.008$ & \cite{Tombesi10} & NO\\
   
   & $0.428\pm0.008$ & \cite{Tombesi10} & NO\\
   
PKS0558-50 & NO & \cite{Tombesi14} & NO\\
   
UGC 3973 & NO & \cite{Tombesi10} & $0.091\pm0.004$\\
   \hline
\end{tabular}
\vspace{2mm}
\tablefoot{Column 1: Source name; Column 2: our results; Column 3: previous studies on the source; Column 4: results in previous work.\\
*An absorption feature at the same energy is detected but interpreted as an Fe H-like edge with $v\sim0.05 \rm c$, following \citealt{Pounds03mrk766}, see Sec. \ref{Sec. 5} for details.
\textdagger A subthreshold feature at similar energies is reported in Tab. \ref{discarded}.
$\ddagger$ A UFO with v$_{\rm out}=0.2 \rm c$ is detected in the same observation in the soft band through RGS spectroscopy \citep{Xu2023}. 
}
\label{Table 6.1}
\end{table}

\end{appendix}

\end{document}